\documentclass[twocolumn,floatfix,showpacs,aps,prd]{revtex4}

   \usepackage{graphicx}
   \usepackage{dcolumn}
   \usepackage{amsmath}
   \usepackage{units}
   \usepackage[abs]{overpic}
   \usepackage{float}
   \usepackage{wrapfig}
   \input{babarsym}

   \newcommand{\Btag}{\ensuremath{\B_\mathrm{tag}}\xspace}
   
   \newcommand{\Bsig}{\ensuremath{\B_\mathrm{sig}}\xspace}
   \newcommand{\btodlnu}{\ensuremath{B \to D^{(*)0}\ell\nu}\xspace}
   \newcommand{\btodmnu}{\ensuremath{B \to D^{(*)0}\mu\nu}\xspace}
   \newcommand{\btodenu}{\ensuremath{B \to D^{(*)0}e\nu}\xspace}
   \newcommand{\btohtl}{\ensuremath{B \to h \tau \ell}\xspace}
   \newcommand{\btoktl}{\ensuremath{B \to K \tau \ell}\xspace}
   \newcommand{\btoptl}{\ensuremath{B \to \pi \tau \ell}\xspace}
   \newcommand{\btoptm}{\ensuremath{B \to \pi \tau \mu}\xspace}
   \newcommand{\btopte}{\ensuremath{B \to \pi \tau e}\xspace}
   \newcommand{\btohte}{\ensuremath{B \to h \tau e}\xspace}
   \newcommand{\btohtm}{\ensuremath{B \to h \tau \mu}\xspace}
   \newcommand{\btoktm}{\ensuremath{B \to K \tau \mu}\xspace}

   \newcommand{\bptokptm}{\ensuremath{B^+ \to K^+ \tau \mu}\xspace}
   \newcommand{\bptokpte}{\ensuremath{B^+ \to K^+ \tau e}\xspace}
   \newcommand{\bptopiptm}{\ensuremath{B^+ \to \pi^+ \tau \mu}\xspace}
   \newcommand{\bptopipte}{\ensuremath{B^+ \to \pi^+ \tau e}\xspace}

   \newcommand{\btokte}{\ensuremath{B \to K \tau e}\xspace}
   \newcommand{\btoktmp}{\ensuremath{B^+ \to K^+ \tau^- \mu^+}\xspace}
   \newcommand{\btoktmm}{\ensuremath{B^+ \to K^+ \tau^+ \mu^-}\xspace}
   \newcommand{\btoktep}{\ensuremath{B^+ \to K^+ \tau^- e^+}\xspace}
   \newcommand{\btoktem}{\ensuremath{B^+ \to K^+ \tau^+ e^-}\xspace}
   \newcommand{\btoptmp}{\ensuremath{B^+ \to \pi^+ \tau^- \mu^+}\xspace}
   \newcommand{\btoptmm}{\ensuremath{B^+ \to \pi^+ \tau^+ \mu^-}\xspace}
   \newcommand{\btoptep}{\ensuremath{B^+ \to \pi^+ \tau^- e^+}\xspace}
   \newcommand{\btoptem}{\ensuremath{B^+ \to \pi^+ \tau^+ e^-}\xspace}
   \newcommand{\smtvs}{\rule[-0.15cm]{0cm}{0.5cm}}
   \newcommand{\bptoktlp}{\ensuremath{B^+ \to K^+ \tau^- \ell^+}\xspace}
   \newcommand{\bptoktlm}{\ensuremath{B^+ \to K^+ \tau^+ \ell^-}\xspace}
   
   \newcommand{\bptoptlm}{\ensuremath{B^+ \to \pi^+ \tau^+ \ell^-}\xspace}
   \newcommand{\bptohtlp}{\ensuremath{B^+ \to h^+ \tau^- \ell^+}\xspace}

   \newcommand{\Dstarzbpar}{\ensuremath{\Dbar^{(*)0}}\xspace}
   \newcommand{\Dstarstarzb}{\ensuremath{\Dbar^{**0}}\xspace}
   \newcommand{\Dstarstarz}{\ensuremath{D^{**0}}\xspace}
   
   \newcommand{\dlnu}{\ensuremath{D^{(*)0}\ell\nu}\xspace}
   \newcommand{\costhr}{\ensuremath{|\cos \theta_{\rm thr}|}\xspace}
   \newcommand{\sumecal}{\ensuremath{\sum E_{\rm cal}}\xspace}
   \newcommand{\tautoenunu}{\ensuremath{\tau \to e \nunub}\xspace}
   \newcommand{\tautomununu}{\ensuremath{\tau \to \mu \nunub}\xspace}
   \newcommand{\tautopinu}{\ensuremath{\tau \to (n\pi^0) \pi \nu}\xspace}

   \newcommand{\BABARPubYear}    {11}
   \newcommand{\BABARPubNumber}  {023}
   \newcommand{\SLACPubNumber} {14951}

   \def\figurebox#1#2#3{%
       \def\arg{#3}%
       \ifx\arg\empty
       {\hfill\vbox{\hsize#2\hrule\hbox to #2{\vrule\hfill\vbox to #1{\hsize#2\vfill}\vrule}\hrule}\hfill}%
       \else
       {\hfill\epsfbox{#3}\hfill}%
       \fi}

   \long\def\inst#1{\par\nobreak\kern 4pt\nobreak
       {\it #1}\par\vskip 10pt plus 3pt minus 3pt}

\begin{document}

   %\linenumbers

   \preprint{\babar-PUB-\BABARPubYear/\BABARPubNumber}
   \preprint{SLAC-PUB-\SLACPubNumber}

   \begin{flushleft}
    SLAC-PUB-14951 \\
    BABAR-PUB-11/023\\
   \end{flushleft}

   \title{
     { \Large \bf \boldmath A search for the decay modes $B^\pm \to h^\pm \tau \ell$ }
   }
    %% author list as of 30-Sep-2011 (379 authors)
%
\author{J.~P.~Lees}
\author{V.~Poireau}
\author{V.~Tisserand}
\affiliation{Laboratoire d'Annecy-le-Vieux de Physique des Particules (LAPP), Universit\'e de Savoie, CNRS/IN2P3,  F-74941 Annecy-Le-Vieux, France}
\author{J.~Garra~Tico}
\author{E.~Grauges}
\affiliation{Universitat de Barcelona, Facultat de Fisica, Departament ECM, E-08028 Barcelona, Spain }
\author{D.~A.~Milanes$^{a}$}
\author{A.~Palano$^{ab}$ }
\author{M.~Pappagallo$^{ab}$ }
\affiliation{INFN Sezione di Bari$^{a}$; Dipartimento di Fisica, Universit\`a di Bari$^{b}$, I-70126 Bari, Italy }
\author{G.~Eigen}
\author{B.~Stugu}
\affiliation{University of Bergen, Institute of Physics, N-5007 Bergen, Norway }
% ...
\collaboration{The \babar\ Collaboration}
\noaffiliation

    \collaboration{The \babar\ Collaboration}

   \begin{abstract}

   We present a search for the lepton flavor violating decay modes $B^{\pm} 
   \to h^{\pm} \tau \ell$ ($h= K,\pi$; $\ell= e,\mu$) using
   the \babar\ data sample, which corresponds to 472 million \BB pairs.
   The search uses events where one $B$ meson is fully reconstructed
   in one of several hadronic final states.
   Using the momenta of the reconstructed $B$, $h$, and $\ell$ candidates,
   we are able to fully determine the $\tau$ four-momentum.
   The resulting $\tau$ candidate mass is our main discriminant against 
   combinatorial background.
   We see no evidence for $B^{\pm}\to h^{\pm} \tau \ell$ decays and
   set a 90\% confidence level upper limit on each branching fraction at the
   level of a few times $10^{-5}$.

   \end{abstract}


   \maketitle

     %%-- Introduction

   %%%%%%%%%%%%%%%%%%%%%%%%%%%%%%%%%%%%%%%%%%%%%%%%%%%%%%%%%%%%%%%%%%%%%%%%%%%%%%%%%%%%%%%%%%%%
   \section{ Introduction }
   \label{sec:intro}

     The standard model (SM) of electroweak interactions does not allow
     charged lepton flavor violation or flavor changing neutral currents (FCNC)
     in tree-level interactions~\cite{SM}.
     Lepton flavor violating decays of $B$ mesons can occur at the
     one-loop level through processes that involve neutrino mixing,
     but these are highly suppressed by powers of $m^2_\nu / m^2_W$~\cite{he}
     and have predicted branching fractions many orders of magnitude
     below the current experimental sensitivity.
     However, in many extensions of the SM, $B$ decays involving
     lepton flavor violation and/or FCNC interactions are greatly
     enhanced~\cite{he,sher-and-yuan,black,fujihara}.
     In some cases, decays involving the second and third
     generations of quarks and leptons are particularly sensitive to
     physics beyond the SM~\cite{sher-and-yuan}.

     Until recent years, experimental information on $B$ decays
     to final states containing $\tau$ leptons has been weak or absent.
     The presence of at least one neutrino from the $\tau$ decay prevents
     direct reconstruction of the $\tau$, making it difficult to
     distinguish $B\to X \tau$ decays from
     the abundant semileptonic $B\to X \ell \nu; \ \ell=e,\mu$ decays.
     The high-luminosity $B$ factory experiments have developed the technique
     of using a fully-reconstructed hadronic $B$ decay (the ``tag'' $B$)
     to determine the three-momentum of the other $B$
     (the ``signal'' $B$) in $\Upsilon(4S) \to \BB$ events,
     which enables the $\tau$ to be indirectly reconstructed.
     This technique assigns all detected tracks
     and neutral objects to either the tag $B$ or the signal $B$.
     Recent applications of this technique by \babar\ are the
     searches for \bptokptm~\cite{babar-ktm},
     $B^0\to\ell^\pm\tau^\mp$~\cite{babar-tl}
     and $B^+ \to \tau^+ \nu$~\cite{babar-tnu}.
     We present an update of our search for \bptokptm~\cite{babar-ktm}
     and the first search for the decays \bptokpte,
     \bptopiptm, and \bptopipte~\cite{chargecon}.

     The signal branching fraction is determined by using
     the ratio of the number of \btohtl ($h=K^\pm,\pi^\pm$) signal candidates to the yield
     of control samples of $B^+ \to \Dbar^{(*)0} \ell^+ \nu; \Dbar^0 \to K^+ \pi^-$ events
     from a fully reconstructed hadronic $B^\pm$ decay sample.
     Continuum background is suppressed for each decay
     channel using a likelihood ratio based on event shape information,
     unassociated calorimeter clusters, and the quality of
     muon identification for channels that have a muon
     in the final state.
     Final signal candidates are selected requiring the
     indirectly reconstructed $\tau$ mass to fall in a narrow window
     around the known $\tau$ mass.
     The yield and estimated background in the $\tau$
     mass signal window are used to estimate and set
     upper limits on the signal branching fractions.
     We followed the principle of a blind analysis, to avoid
     experimenter's bias, by not revealing the number of
     events in the signal window until after all analysis procedures
     were decided.

   %%%%%%%%%%%%%%%%%%%%%%%%%%%%%%%%%%%%%%%%%%%%%%%%%%%%%%%%%%%%%%%%%%%%%%%%%%%%%%%%%%%%%%%%%%%%
   \section{ Data sample and Detector description }
   \label{sec:datadetect}

     We use a data sample of 472 million \BB pairs
     in 429~fb$^{-1}$ of integrated luminosity, delivered by the PEP-II
     asymmetric-energy $e^+e^-$ collider and recorded by the \babar\ experiment
     at the SLAC National Accelerator Laboratory.
     This corresponds to the entire $\Upsilon(4S)$ data sample.

     The \babar\ experiment is described in detail elsewhere~\cite{ref:babar}.
     Trajectories of charged particles are reconstructed by a 
     double-sided, five-layer silicon vertex tracker (SVT) and
     a 40-layer drift chamber (DCH).
     The SVT provides precision measurements for vertex reconstruction and
     stand-alone tracking for very low momentum tracks, with transverse
     momentum less than 120\mevc.
     The tracking system is inside a 1.5 T superconducting solenoid.
     Both the SVT and the DCH provide specific ionization ($dE/dx$)
     measurements that are used in particle identification (PID).
     Just beyond the radius of the DCH lies an array of fused silica bars which
     are part of the detector of internally reflected Cherenkov radiation (DIRC).
     The DIRC provides excellent charged-hadron PID.
     A CsI(Tl) crystal electromagnetic calorimeter (EMC)
     is used to reconstruct photons and identify electrons.
     The minimum EMC cluster energy used in this analysis is 30 MeV.
     The iron of the flux return for the solenoid is instrumented (IFR) with
     resistive plate chambers and limited streamer tubes, which are
     used in the identification of muons.

     Monte Carlo (MC) simulated samples for our \btohtl\ signals and for
     all relevant SM processes are generated
     with EvtGen~\cite{ref:evtgen}.
     We model the \babar\ detector response using \geant~\cite{ref:geant}.
     The \btohtl\ decays are generated using a uniform three-body phase space
     model and the background MC sample combines SM processes:
     $\epem\ \to \Upsilon(4S) \to \BB$, $\epem\ \to q\bar{q}$ ($q = u,d,s,c$), and
     $\epem\ \to \tau^+\tau^-$.
     The number of simulated Monte Carlo events corresponds to integrated
     luminosities equivalent to three times the data for \BB events and
     two times the data for the continuum processes.
     Each Monte Carlo sample is reweighted to correspond to an integrated
     luminosity equivalent to the data.

     The data and MC samples in this analysis are processed and generated
     with consistent database conditions determined from the detector response
     and analyzed using \babar\ analysis software release tools.

     %%%%%%%%%%%%%%%%%%%%%%%%%%%%%%%%%%%%%%%%%%%%%%%%%%%%%%%%%%%%%%%%%%%%%%%%%%%%%%%%%%%%%%%%%%%%

     \section{ Event reconstruction }
     \label{sec:eventreco}

     In each event, we require a fully reconstructed hadronic $B^\pm$ decay,
     which we refer to as the ``tag'' $B$ meson candidate or \Btag.
     We then search for the signal \btohtl\ decay in the rest of
     the event, which we refer to as the ``signal'' $B$ meson candidate or \Bsig.
     The notation \btohtl\ refers to one of the following eight final states that
     we consider, where the primary hadron $h$ is a $K$ or $\pi$
     and the primary lepton $\ell$ is a $\mu$ or $e$:
     \btoktmp, \btoktmm, \btoktep, \btoktem, \btoptmp, \btoptmm, \btoptep, and \btoptem.
     In all cases, we require that the $\tau$ decays to a ``one-prong'' final state
     (\tautoenunu, \tautomununu, and \tautopinu with $n \ge 0$).
     The branching fraction for $\tau$ decays to a one-prong final state is 85\%.

     The $\Upsilon(4S) \to B^+B^-$ decay requires
     the \Bsig three-momentum to be opposite from that of the \Btag ($-\vec p_{\rm tag}$)
     and the \Bsig energy to be equal to the beam energy ($E_{\rm beam}$)
     in the \epem\ center-of-mass (CM) reference frame~\cite{cmframe}.
     These constraints allow us to reconstruct the $\tau$ indirectly using
     \begin{eqnarray*}
       \vec p_\tau & = & -\vec p_{\rm tag} - \vec p_h - \vec p_\ell , \\
       E_\tau & = & E_{\rm beam} - E_h - E_\ell , \\
       m_\tau & = & \sqrt{ E_\tau^2 - |\vec{p}_\tau|^2 } ,
     \end{eqnarray*}
     where ($E_\tau$, $\vec p_\tau$), ($E_h$, $\vec p_h$), and ($E_\ell$, $\vec p_\ell$)
     are the corresponding four-momenta of the reconstructed signal objects.
     The indirectly reconstructed $\tau$ mass ($m_\tau$) peaks sharply at the true
     $\tau$ mass in \btohtl\ signal events and has a very broad distribution for
     combinatorial background events.
     To avoid experimental bias, we did not look at events in the data with
     $m_\tau$ within $\pm 175$\mevcc of the nominal $\tau$ mass until all
     analysis procedures were established.

     %%-----------------------------------------------------------------
     \subsection{ \boldmath Tag $B$ reconstruction }
     \label{subsec:tagb}

     The \Btag is fully reconstructed in one of many final states~\cite{ref:Brecoil}
     of the form $B^- \to D^{(*)0}X^-$.
     The notation $D^{(*)0}$ refers to either a $D^0$ or a $D^{*0}$
     which decays to either $D^0\gamma$ or $D^0\pi^0$.
     The $D^0$ is reconstructed in the $K^-\pi^+$, $K^-\pi^+\pi^-\pi^+$, $K^-\pi^+\pi^0$,
     and $K_S^0\pi^+\pi^-$ channels, with $K^0_S \to \pi^+\pi^-$ and $\pi^0\to \gamma\gamma$.
     The $X^-$ represents a system of charged and neutral
     hadrons composed of $n_1 \pi^\pm$, $n_2 K^\pm$, $n_3 K_S^0$, and
     $n_4 \pi^0$; subject to the constraints $n_1 + n_2 \leq 5$, $n_3 \leq 2$, $n_4 \leq 2$,
     and total charge $-1$.

     Each distinct \Btag decay mode has an associated {\it a priori} purity,
     defined as the number of peaking events divided by the number of peaking
     plus combinatorial events, where peaking and combinatorial yields are obtained
     from fits to $\mes \equiv \sqrt{ E^2_{\rm beam} - |\vec{p}_{\rm tag}|^2 }$
     distributions for each distinct \Btag decay mode.
     We only consider \Btag decay modes with a purity greater than 10\%
     and choose the \Btag candidate with the highest purity in the event.
     If there is more than one \Btag candidate with the same purity, we
     choose the one with reconstructed energy closest to the beam
     energy.
     The \Btag candidate must have $\mes > 5.27$\gevcc
     and $E_{\rm tag}$ within three standard deviations of $E_{\rm beam}$.
     A charged \Btag candidate is properly reconstructed in approximately 0.25\% of
     all \BB events.

     %%-----------------------------------------------------------------
     \subsection{ Particle identification }
     \label{subsec:pid}

     PID algorithms are used to identify kaons, pions, protons, muons, and electrons.
   We use an error correcting output code (ECOC) algorithm~\cite{ecoc} with
   36 input variables to identify electrons, pions, and protons.
   The ECOC combines multiple bootstrap aggregated decision tree
   binary classifiers trained to separate $e, \pi, K,$ and $p$.
   The most important inputs for electron identification
   are the EMC energy divided by the track momentum,
   several EMC shower shape variables, and the deviation from the expected
   values divided by the measurement uncertainties of the
   Cherenkov angle and of the $dE/dx$ for the $e, \pi, K,$ and $p$ hypotheses.
   Neutral clusters in the EMC that are consistent with bremsstrahlung
   radiation are used to correct the momentum and energy of electron
   candidates.
   A $\gamma$ candidate from an $e^\pm$ track is consistent
   with bremsstrahlung radiation if the corresponding
   three-momenta are within $|\Delta \theta| < 35$\mrad
   and $|\Delta \phi| < 50$\mrad, with respect to the polar
   and azimuthal angles of the beam axis.

   Muons and kaons are identified using a bagging decision trees~\cite{narsky}
   algorithm with 30 (36) input variables for the muon (kaon) selection.
   For muons, the most important input variables are
   the number and position of the hits in the IFR, the
   difference between the expected and measured DCH $dE/dx$
   for the muon hypothesis, and the energy deposited in the EMC.
   For kaons, the most important variables are the kaon and pion likelihoods
   based on the measured Cherenkov angle in the DIRC and the
   difference between the expected and measured $dE/dx$
   for the kaon hypothesis.

   We define several quality levels of particle identification for use
   in the analysis.  The ``loose'' levels have higher efficiency but also
   higher misidentification probabilities.  The ``tight'' levels have lower
   misidentification probabilities and efficiencies.
     Table~\ref{tab:pid} summarizes the selection efficiency and misidentification probabilities of
     the PID selection algorithms used.
     A ``very loose'' (VL) $K$-PID algorithm is used for identifying the primary $K$
     in \btoktl, while a ``very tight'' (VT) $K$-PID algorithm, with lower efficiency
     but much smaller misidentification probability, is used to
     reject \Bsig candidates where a non-kaon track passes the the VT $K$-PID criteria.
     Four quality levels of $\mu$-PID are used.
     In order of decreasing efficiency and misidentification probability, they are Very Loose (VL), Loose (L),
     Tight (T), and Very Tight (VT).

     \begin{table}
       \begin{center}
         \caption{PID efficiencies and misidentification probabilities for the algorithms
                  used in the analysis. The values are approximate and representative
                  only for the laboratory frame momentum ($p_{\rm lab}$) specified (when given).
                  More than one algorithm is used for kaons and muons.
                  The abbreviations VL, L, T, and VT stand for selection quality levels
                  Very Loose, Loose, Tight, and Very Tight.
                  }
         \label{tab:pid}
         \begin{tabular}{ccc}
           \hline \hline
            Type & Efficiency & misidentification probability  \smtvs \\
            \hline\hline
            $K$-VL   &  $>95\%$ &  $<6$\%  for $\pi$ and $\mu$ with $p_{\rm lab}<3.5$\gevc  \smtvs \\
               \hline
            $K$-VT   &  $>85\%$ &  $\approx 1$\%  for $\pi$ and $\mu$ with $p_{\rm lab}<3.5$\gevc  \smtvs \\
            \hline
            $\pi$    &  $>98\%$ & $<20$\% for $K$ \smtvs \\
            \hline
            $p$      &  $\approx80$\%  & $<0.5$\% for $K,\pi, \mu, e$ \smtvs \\
            \hline
            $\mu$-VL &  $\approx90$\%  &  $<15$\% for $\pi$ with $p_{\rm lab}<1.25$\gevc, \smtvs \\
                     &                 &   $<4$\% for $\pi$ with $p_{\rm lab}>1.25$\gevc  \\
                \hline
            $\mu$-L  &  $\approx80$\%  &   $<5$\% for $\pi$ with $p_{\rm lab}<1.25$\gevc, \smtvs \\
                     &                 &   $<2$\% for $\pi$ with $p_{\rm lab}>1.25$\gevc   \\
                \hline
            $\mu$-T  &  $\approx75$\%  &   $<3$\% for $\pi$ with $p_{\rm lab}<1.25$\gevc, \smtvs \\
                     &                 &   $\approx 1$\% for $\pi$ with $p_{\rm lab}>1.25$\gevc  \\
                \hline
            $\mu$-VT &  $\approx70$\%  &   $<2$\% for $\pi$ with $p_{\rm lab}<1.25$\gevc, \smtvs \\
                     &                 &   $<1$\% for $\pi$ with $p_{\rm lab}>1.25$\gevc  \\
            \hline
            $e$      &  95\%           &  $<0.2$\% for $\pi,K,p$ \smtvs \\
            \hline \hline
         \end{tabular}
       \end{center}
     \end{table}

     %%-----------------------------------------------------------------
     \subsection{ \boldmath Signal $B$ reconstruction }
     \label{subsec:sigb}

     The eight \btohtl\ decay modes are independently analyzed.
     Tracks for the signal $B$ reconstruction must satisfy the following
     criteria: the distance of closest approach (DOCA) to the beam
     axis in the transverse plane must be less than 1.5~cm; the
     $z$ position of the DOCA point must be less than 2.5~cm from
     the primary vertex of the event; the transverse momentum must
     be $>50$~\mevc; and the momentum must be $<10$~\gevc.
     After selecting the best \Btag candidate, we require exactly
     three tracks satisfying the above criteria remain in the event (excluding the \Btag daughters)
     and that the sum of the charges of these tracks be the opposite of the \Btag
     candidate charge.
     We refer to these three tracks as the \Bsig daughters.

     We require the primary hadron, which is the $h$ in \btohtl, to be
     one of the two \Bsig daughters with the same charge as the \Bsig candidate.
     The primary hadron must pass the $K$-VL-PID criteria for the \btoktl\ modes
     and the $\pi$-PID criteria for the \btoptl\ modes.
     For the \btoktl\ modes, if both of the \Bsig daughters with the same charge meet the minimal $K$-PID
     criteria, the one with the highest $K$-PID quality level is selected as the primary $K^\pm$.
     If they have the same $K$-PID quality level, we choose the one with the
     lower momentum as the primary $K^\pm$.
     For the \btoptl\ modes, if both \Bsig daughters with the same charge meet the
     $\pi$-PID criteria, we choose the one that gives $m_\tau$ closest to the true $\tau$
     mass.
     This algorithm does not produce an artificial peak in the signal window
     of the background $m_\tau$ distribution.
     Once the primary hadron candidate has been assigned, the $\tau$ daughter and
     primary lepton are uniquely defined for a given \btohtl\ mode from the remaining
     two \Bsig daughters based on their electric charge.

     The primary lepton, which is the $\ell$ in \btohtl, must pass
     either the $e$-PID or the loosest $\mu$-PID criteria ($\mu$-VL).
     We remove events where any of the three \Bsig daughters passes
     the $p$-PID criteria, or where any of the three \Bsig daughters
     passes the $K$-VT-PID criteria, with the exception of the $K^\pm$
     in \btoktl.

     By requiring exactly three \Bsig daughters, we are restricting the selection
     to one-prong $\tau$ decays.
     For each of the eight \btohtl\ modes, we divide the selection into three
     $\tau$ decay channels: electron, muon, and pion.
     From now on, we use ``modes'' to refer to types of \btohtl\ decays
     and ``channels'' to refer to types of $\tau$ decays.
     The three $\tau$ decay channels are analyzed in parallel, with different
     background rejection criteria applied.
     If the $\tau$ daughter satisfies the $e$-PID criteria, the event is assigned
     to the electron channel.
     If the $\tau$ daughter does not satisfy the $e$-PID, but does satisfy the
     $\mu$-VL-PID criteria, the event is assigned to the muon channel.
     If the $\tau$ daughter passes neither the $e$-PID or the $\mu$-VL-PID, the event
     is assigned to the pion channel.
     This ensures that an event does not get double counted and categorized into
     another $\tau$ decay channel for a given \btohtl\ mode.

     Background events with a $B \to h (c\bar c) ; \ (c\bar c) \to \ell^+\ell^-$
     decay can pass our signal selection criteria.
     We remove events in the electron (muon) and pion $\tau$ decay channels
     of the \btohte\ (\btohtm) modes if the invariant mass of the primary
     lepton and $\tau$ daughter, $m_{\ell\ell}$, is consistent
     with a dilepton charmonium decay: $3.03<m_{\ell\ell}<3.14$\gevcc
     for the $J/\psi$ or $3.60<m_{\ell\ell}<3.75$\gevcc for the $\psi(2S)$.
     The core dilepton invariant mass resolution for these charmonium
     decays is on the order of 12~\mevcc.  These charmonium vetos
     effectively remove the charmonium background at a minimal cost in
     signal efficiency.
     We also require $m_{\ell\ell}>0.1$\gevcc for \btohte\ candidates in the electron
     and pion channels to remove candidates where the primary electron and the $\tau$
     daughter are consistent with originating from a photon conversion.

   %%============
     \subsection{ \boldmath \BB background and the $m(K\pi)$ invariant mass requirement }

     After the selection described above, the dominant background is due to \BB
     events, where the \Btag is properly reconstructed.
     However, the largest background source differs depending on the charge of
     the primary lepton relative to the charge of the \Bsig candidate.

     When the primary lepton charge is the same as the \Bsig charge,
     such as a \bptoktlp candidate, the dominant background comes from
     semileptonic $B$ decays, such as $B^+ \to \Dstarzbpar \ell^+ \nu; \ \Dzb \to K^+ X^-$,
     where $X^-$ contains a $\pi^-$, $e^-$, or $\mu^-$ and perhaps other
     charged and/or neutral daughters that are not reconstructed.
     For example, the final state tracks $K^+\pi^-\ell^+$ are identical for
     this background with $\Dzb \to K^+ \pi^-$ and the \bptoktlp signal decay
     with $\tau^- \to \pi^- \nu_\tau$.
     On the other hand, when the primary lepton charge is opposite to the \Bsig
     charge, such as for a \bptoktlm candidate, the dominant background comes from
     semileptonic $D$ decays, such as
     $B^+ \to \Dstarzbpar X^+; \ \Dzb \to K^+ \ell^- \bar \nu_\ell$.

\begin{figure*}
      \begin{center}
         \begin{overpic}[width=0.5\linewidth,unit=1mm]%
            {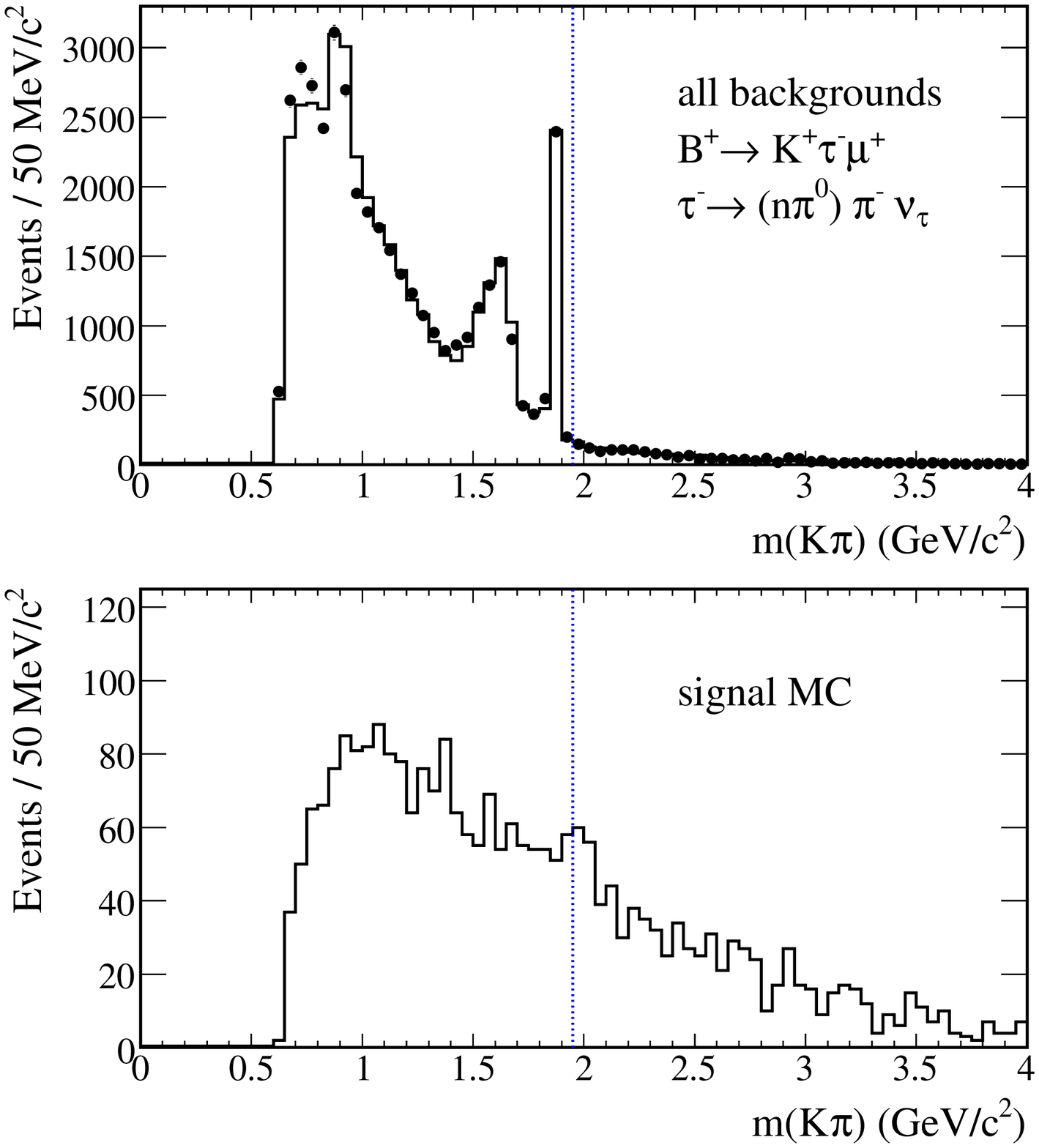}
            \put(17,90){(a)}\put(17,42){(b)}
         \end{overpic}%
         \begin{overpic}[width=0.5\linewidth,unit=1mm]%
            {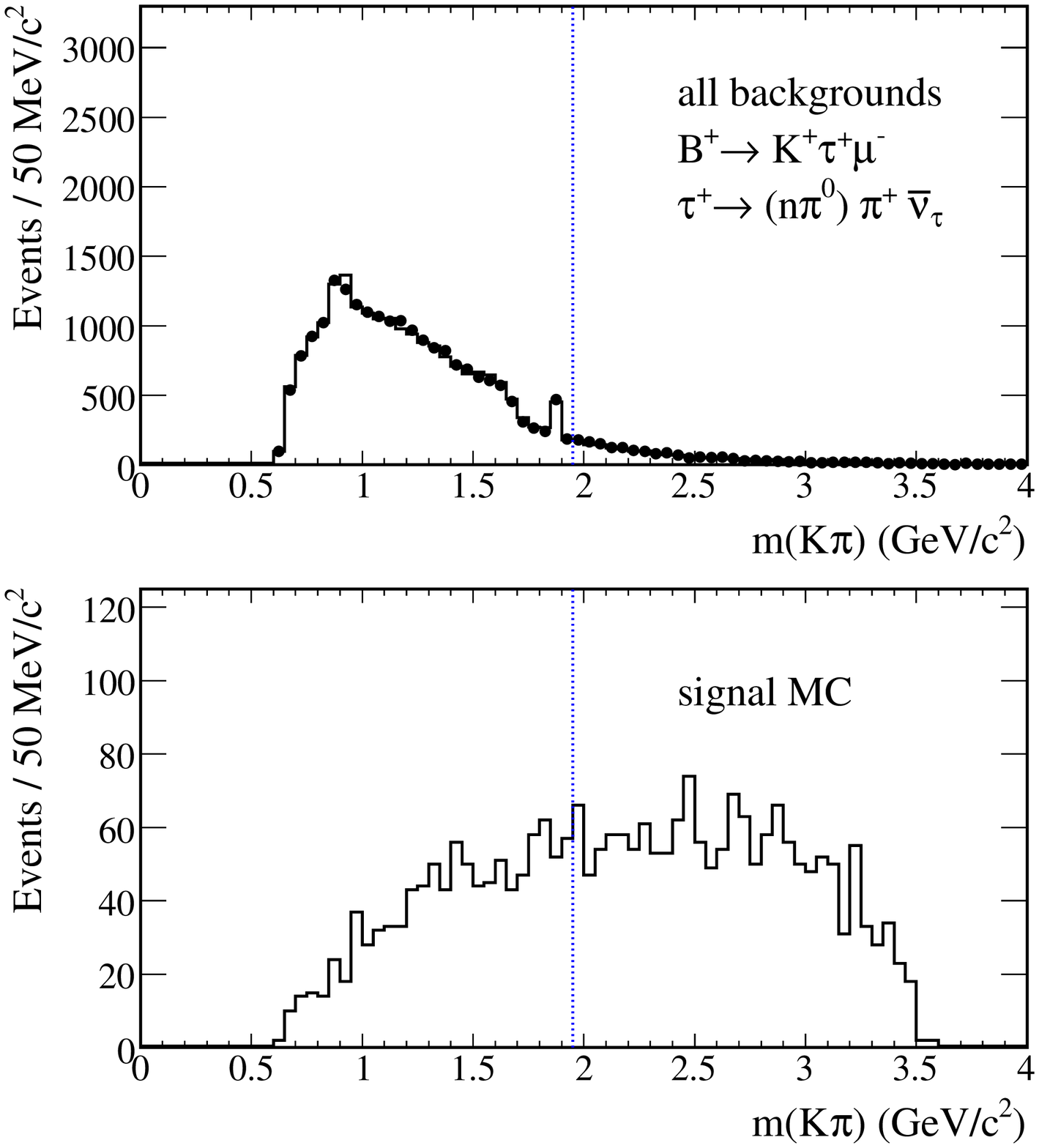}
            \put(17,90){(c)}\put(17,42){(d)}
         \end{overpic}
         \caption{
           Distributions of $m(K\pi)$ for the 
           (a, b) \btoktmp;
           $\tau^- \to (n\pi^0)\pi^-\nu_\tau$
           and 
           (c, d) \btoktmm;
           $\tau^+ \to (n\pi^0)\pi^+\bar\nu_\tau$
           channels.
           The top row shows the data (points) compared with the
           background MC (solid line).  The area of the background MC distribution has been
           normalized to the area of the data distribution.
           The bottom row shows the (b) \btoktmp and (d) \btoktmm signal MC.
           The normalization of the bottom row is arbitrary.
           The dotted vertical line is at 1.95\gevcc, which is the minimum
           allowed value of $m(K\pi)$ for the signal selection.
           The peak in the top row just below 1.95\gevcc is from
           $\Dzb \to K^+ \pi^-$ decays.
         }
         \label{fig:mkpi}
      \end{center}
   \end{figure*}

     To reduce these backgrounds, we reject \Bsig candidates where
     two of the \Bsig daughters are kinematically compatible with
     originating from a charm decay, as described below.
     For the four \btoktl\ modes, we define the variable $m(K\pi)$ as the invariant mass
     of the primary $K$ and the \Bsig daughter that has opposite charge to this $K$.
     In computing $m(K\pi)$, the non-$K$ track is assumed to be a pion.
     Distributions of $m(K\pi)$ for the background and signal MC are shown
     in Fig.~\ref{fig:mkpi} for \btoktm.
     For the four \btoptl\ modes, we define $m(K\pi)$ by combining two \Bsig daughters
     that have opposite charge.
     Of the two \Bsig daughters with the same charge as the \Bsig candidate, we choose
     the one with the highest $K$-PID quality level.
     We assume that the kaon is one of the \Bsig daughters with the same charge
     as the \Bsig candidate and the pion is the \Bsig daughter with the opposite
     charge as the \Bsig candidate.
     If the two \Bsig daughters with the same charge as the \Bsig candidate have the
     same $K$-PID quality level, we use the daughter with higher momentum as the
     kaon in the $m(K\pi)$ calculation.

     We require $m(K\pi)>1.95$\gevcc.  This rejects between 97\% and 99\%
     of the background while retaining between 32\% and 37\% of the signal
     for the \bptohtlp modes.
     For the \bptoptlm modes, the $m(K\pi)$ requirement rejects 85\% and 89\% of the
     $\pi^+ \tau^+ \mu^-$ and $\pi^+ \tau^+ e^-$ background while retaining
     72\% and 65\% of the signal, respectively.
     For the \bptoktlm modes, the $m(K\pi)$ requirement rejects 92\% and 96\%
     of the $K^+\tau^+\mu^-$ and $K^+\tau^+e^-$ background while retaining 63\%
     and 62\% of the signal, respectively.

   %
   %----------------------------------------------------------
   %----------------------------------------------------------
   %

  %%===============================================================================
     \section{ \boldmath \btodlnu control sample }

     We select a control sample of semileptonic $B$ decays of the
     form $B^+ \to \Dstarzbpar \ell^+ \nu; \ \Dzb \to K^+ \pi^-$ by requiring
     $m(K\pi)$ to be near the \Dz mass, $1.845 < m(K\pi) < 1.885$\gevcc.
     The \dlnu\ control sample has a negligible amount of combinatorial background.
     In our search for \btohtl, we normalize the \btohtl\ branching fraction by
     using the measured \dlnu\ yield taken from the control sample.
     We determine the relative amounts of $B$ mesons that decay to
     $\Dzb$, $\Dstarzb$, and higher resonances ($\Dstarstarzb$)
     using the reconstructed CM energy difference
     \begin{eqnarray*}
       E_\nu & = & p_\nu = \left| - \vec p_{\rm tag} - \vec p_K - \vec p_\pi - \vec p_\ell \right| , \\
       \Delta E_{D\ell\nu} & = & E_K + E_\pi + E_\ell + E_\nu - E_{\rm beam} .
     \end{eqnarray*}
     For $B^+\to \Dzb \ell^+ \nu$ decays, $\Delta E_{D\ell\nu}$ is centered at zero.
     The missing neutral particles from $\Dstarzb$ and $\Dstarstarzb$ decays
     shift  $\Delta E_{D\ell\nu}$ in the negative direction.

     The expected observed yields of $D\ell\nu$ and $h\tau\ell$ as functions
     of their branching fractions are given by
     \begin{eqnarray}
       \label{eqn:ndlnu}
       N_{D\ell\nu}  & = & N_0 \, {\cal B}_{D\ell\nu} \, \epsilon^{D\ell\nu}_{\rm tag}  \, \epsilon_{D\ell\nu},  \\
       \label{eqn:nhtaul}
       N_{h\tau\ell} & = & N_0 \, {\cal B}_{h\tau\ell} \, \epsilon^{h\tau\ell}_{\rm tag} \, \epsilon_{h\tau\ell},
     \end{eqnarray}
     where $N_0$ is the number of \BB events,
     ${\cal B}_{D\ell\nu}$ (${\cal B}_{h\tau\ell}$) is the branching fraction
     for $B \to D\ell\nu$ ($B \to h\tau\ell$),
     $\epsilon^{D\ell\nu}_{\rm tag}$  ($\epsilon^{h\tau\ell}_{\rm tag}$)
     is the \Btag reconstruction efficiency in \BB events that contain
     a $D\ell\nu$ ($h\tau\ell$) decay on the signal side,
     $\epsilon_{D\ell\nu}$ ($\epsilon_{h\tau\ell}$) is the signal-side
     reconstruction efficiency for $D\ell\nu$ ($h\tau\ell$),
     and the symbol $D\ell\nu$ represents either
     $B^+ \to \Dstarzb \ell^+ \nu$ or
     $B^+ \to \Dzb \ell^+ \nu$.
     Solving for the expected $h\tau\ell$ event yield gives
     \begin{equation}
         N_{h\tau\ell} = {\cal B}_{h\tau\ell} \, \epsilon_{h\tau\ell} \, S_0,
     \end{equation}
     where we have defined a common factor
     \begin{equation}
       S_0 \ = \ \frac{N_{D\ell\nu}}{{\cal B}_{D\ell\nu} \, \epsilon_{D\ell\nu} } \left( \frac{\epsilon^{h\tau\ell}_{\rm tag}}{\epsilon^{D\ell\nu}_{\rm tag}} \right).
     \end{equation}
     Table~\ref{tab:tagside-eff-ratio} gives the tag-side efficiency ratios
     determined from MC samples.
     We find the ratios to be close to one, indicating that the signal-side decay
     does not strongly influence the tag-side reconstruction efficiency,
     and does not depend on the primary lepton or hadron flavor.
     %

     %%---------------------
     \begin{table}
       \begin{center}
         \caption{Tag-side reconstruction efficiency ratios determined from MC samples.
                  The uncertainty includes both statistical and systematic sources.
                  }
         \label{tab:tagside-eff-ratio}
         \begin{tabular}{ccc}
           \hline \hline
           Efficiency Ratio & $\mu$ modes & $e$ modes \\
           \hline\hline
           $\epsilon^{K\tau\ell}_{\rm tag} / \epsilon^{D\ell\nu}_{\rm tag}$   & \ \ \ \ $0.96 \pm 0.05$ \ \ \ \  & $0.98 \pm 0.07$  \\
           $\epsilon^{\pi\tau\ell}_{\rm tag} / \epsilon^{D\ell\nu}_{\rm tag}$ & \ \ \ \ $0.95 \pm 0.04$ \ \ \ \  & $0.97 \pm 0.06$ \\
            \hline \hline
         \end{tabular}
       \end{center}
     \end{table}
     %%---------------------

\begin{figure}[h]
      \begin{center}
         \begin{overpic}[width=0.9\linewidth,unit=1mm]%
            {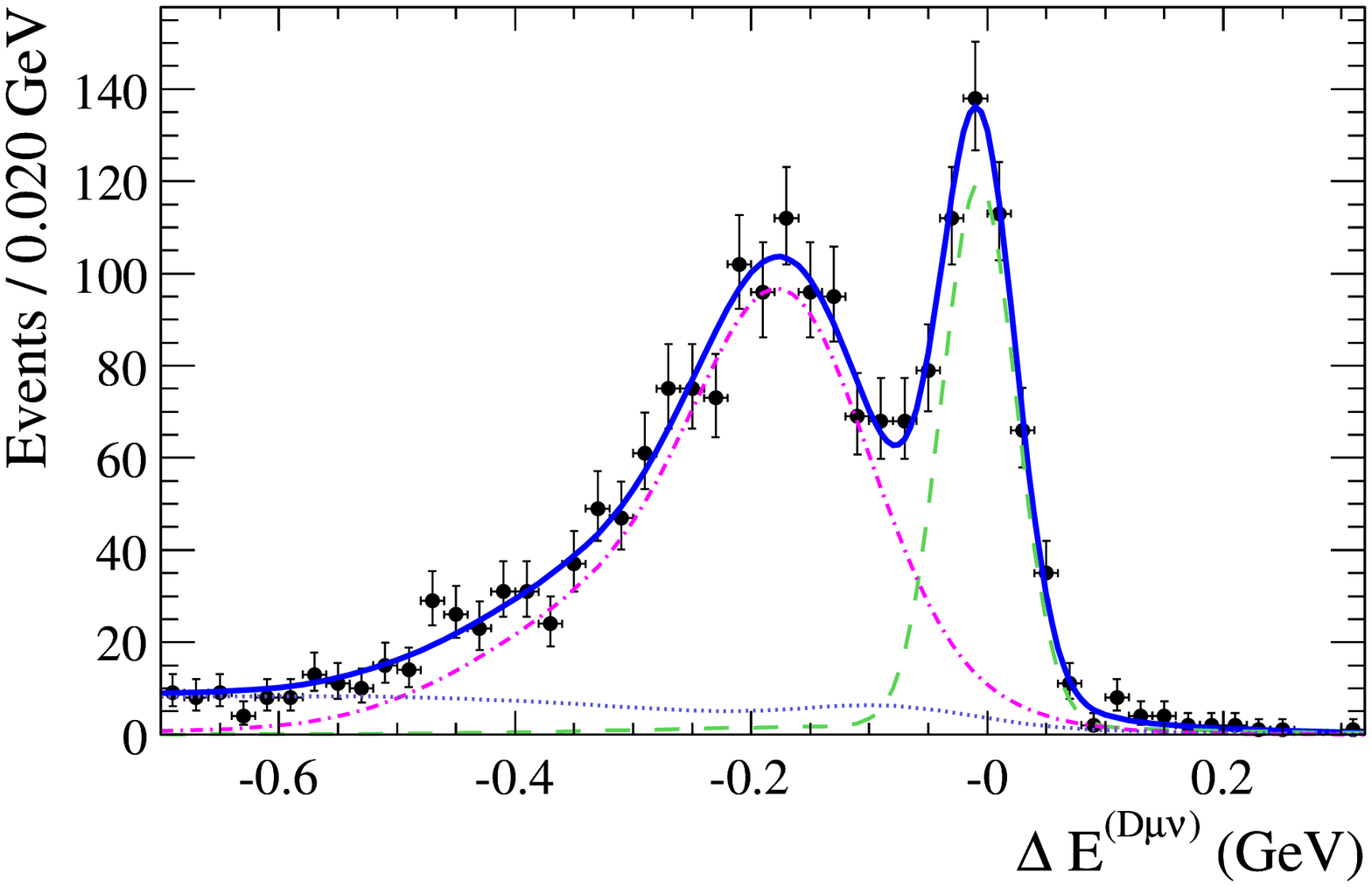}
            \put(20,40){(a)}
         \end{overpic} \\ . \\
         \begin{overpic}[width=0.9\linewidth,unit=1mm]%
            {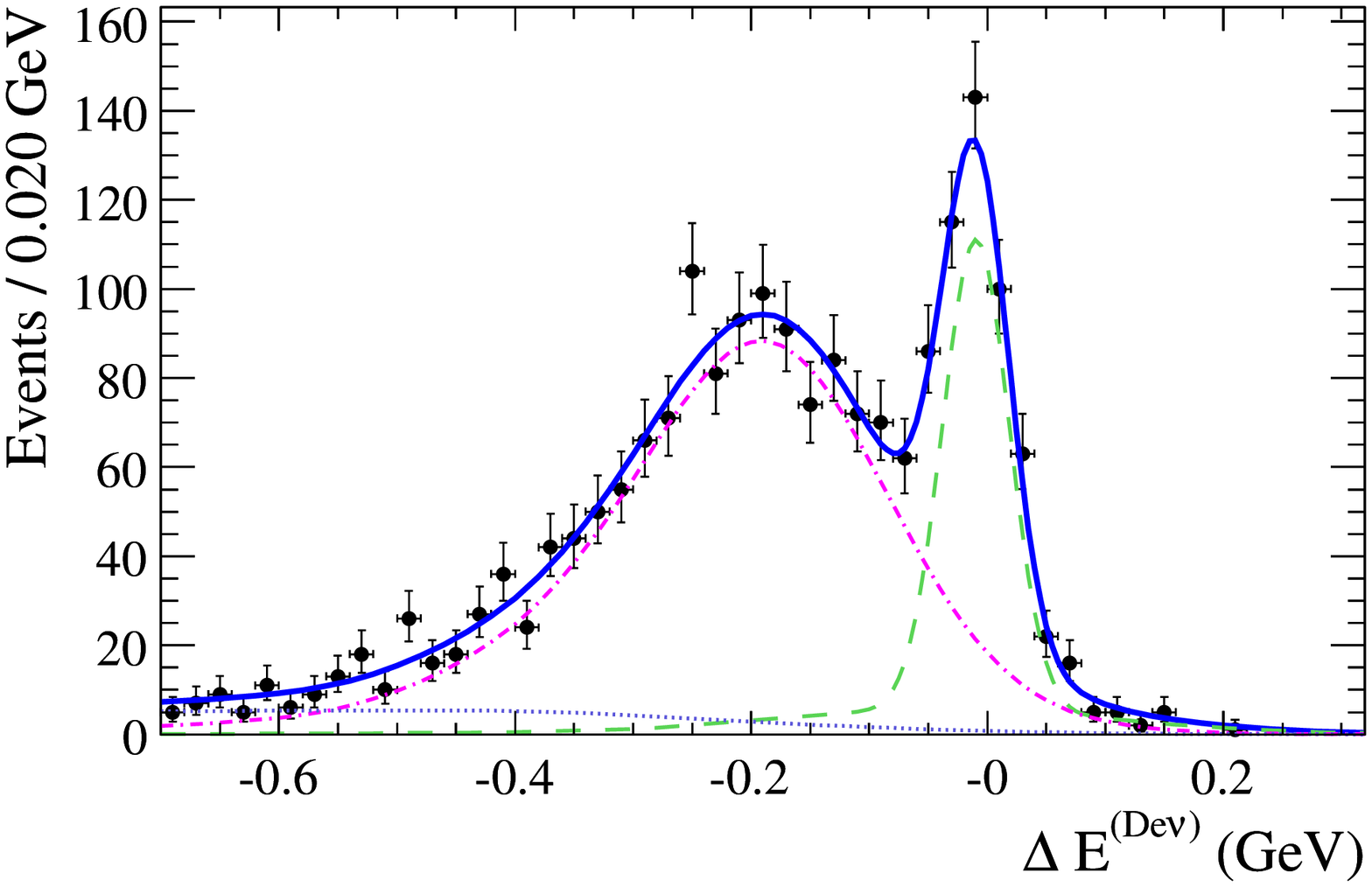}
            \put(20,40){(b)}
         \end{overpic}
         \caption{
            (color online) Distributions of the three-component $\Delta E_{D\ell\nu}$ unbinned
            maximum likelihood fits of the data for the (a) $B\to D^{(*)0} \mu \nu$ and (b)
            $B\to D^{(*)0} e \nu$ control samples.
            In each plot, the points represent the data,
            the solid blue curve is the sum of all PDFs,
            the long-dashed green curve is the \Dz component,
            the dot-dashed purple curve is the \Dstarz component,
            and the dotted blue curve is the \Dstarstarz component, which also
            includes any residual combinatorial background.
         }
         \label{fig:defit}
      \end{center}
   \end{figure}

     Figure~\ref{fig:defit} shows the results of unbinned maximum likelihood fits
     of the $\Delta E_{D\ell\nu}$ distributions for the
     $B^+ \to \Dstarzbpar \mu^+ \nu$ and
     $B^+ \to \Dstarzbpar e^+ \nu$ control samples.
     The fits have independent $\Dzb$, $\Dstarzb$, and $\Dstarstarzb$ components.
     Any residual combinatorial background is included in the  $\Dstarstarzb$ component.
     The $\Dzb$ and $\Dstarzb$ component probability density functions (PDFs)
     are each modeled with the sum
     of a Gaussian and a Crystal Ball function~\cite{cb-function}.
     The $\Dstarstarzb$ component PDF is the sum of a Gaussian and a
     bifurcated Gaussian, which has different width parameters above
     and below the mean.
     The overall normalization of each component,
     the core Gaussian mean and width of the $\Dzb$ component,
     and the relative fraction of the Crystal Ball function within
     the $\Dstarzb$ component are all parameters of the likelihood
     that are varied in its maximization.

     The results of the $\Delta E_{D\ell\nu}$ maximum likelihood fits
     and $S_0$ calculations are given
     in Table~\ref{tab:dlnu-sensitivity-results}.
     We use the following branching fractions~\cite{pdg} in the calculation
     of $S_0$:
     ${\cal B}(B^- \to \Dz \ell^- \bar{\nu}) = (2.23 \pm 0.11)$\%,
     ${\cal B}(B^- \to \Dstarz \ell^- \bar{\nu}) = (5.68 \pm 0.19)$\%, and
     ${\cal B}(\Dz \to K^- \pi^+) = (3.87 \pm 0.05)$\%.
     The four determinations of $S_0$ are all consistent with each other,
     as expected.

   %----------------------------------------------------------
   %----------------------------------------------------------

     %%---------------------
     \begin{table}[ht]
       \begin{center}
         \caption{ Results of the $\Delta E_{D\ell\nu}$ maximum likelihood fits
                   and $S_0$ calculations.  The uncertainties on $N_{D\ell\nu}$
                   and $\epsilon_{D\ell\nu}$ are statistical.
                   The efficiency  $\epsilon_{D\ell\nu}$ is determined from
                   a Monte Carlo sample.
                   The uncertainty on $S_0$ includes the uncertainties on
                   the $B$ and $D$ branching fractions.
                  }
         \label{tab:dlnu-sensitivity-results}
         \begin{tabular}{cccc}
           \hline \hline
           $D\ell\nu$ mode &  $N_{D\ell\nu}$  &  $\epsilon_{D\ell\nu}$  &  $S_0$ \\
           \hline\hline
           $\Dz     \mu \nu$ &  $513 \pm 38$ & ($47.8 \pm 0.9$)\% & $(12.0 \pm 1.2)\times 10^5$ \\
           $\Dstarz \mu \nu$ & \ \ $1234 \pm 49$ \ \ & \ \ ($50.8 \pm 0.5$)\% \ \ & $(10.7 \pm 0.8)\times 10^5$ \\
           \hline
           $\Dz     e   \nu$ &  $484 \pm 46$ & ($48.2 \pm 0.9$)\% & $(11.4 \pm 1.5)\times 10^5$ \\
           $\Dstarz e   \nu$ & $1368 \pm 58$ & ($52.2 \pm 0.5$)\% & $(11.7 \pm 1.1)\times 10^5$ \\
            \hline \hline
         \end{tabular}
       \end{center}
     \end{table}
     %%---------------------

   %%=====================================================================================
     \section{ Continuum background rejection }

     After the $m(K\pi) > 1.95$\gevcc requirement, the \BB background is highly suppressed.
     The remaining background is dominated by continuum quark-pair production
     $(e^+e^- \to q\bar q;\ q=u,d,s,c)$.
     We combine the variables described in this section in a likelihood ratio
     \begin{equation}
       L_R \ = \ \frac{ \prod_i \, P_s(x_i) }{ \prod_i \, P_s(x_i)  +  \prod_i \, P_b(x_i) }
     \end{equation}
     where $x_i$ is one of a set of variables that discriminate against background,
     and $P_s(x_i)$ ($P_b(x_i)$) is the PDF for variable $x_i$ in
     signal (background) events.

     The variables used in the $L_R$ calculation are:
     \begin{itemize}
       \item {\boldmath \costhr } the absolute value of the
          cosine of the angle $\theta_{\rm thr}$ between the \Btag thrust axis
          and the thrust axis of the remainder of the event ($\equiv$\Bsig);
          the thrust axis is defined as the direction $\hat a$ which maximizes
          $\sum_j \, \hat a \cdot \vec p_j$, where $j$ represents all particles
          assigned to a particular $B$ candidate,
       \item {\boldmath \sumecal } the scalar sum of all EMC neutral cluster
          energy that is not associated with the \Btag candidate or bremsstrahlung
          radiation from any $e$ candidates, where the threshold cluster energy
          is 100~MeV (50~MeV) in the forward (barrel) region of the detector,
       \item {\bf \boldmath primary $\mu$-PID quality level}, where, for the \btohtm\ modes,
          we include the highest quality level (VL, L, T, VT) of the primary $\mu$ candidate,
          and
       \item {\bf \boldmath secondary $\mu$-PID quality level}, where we
          include the highest quality level (VL, L, T, VT)
          of the $\tau$-daughter $\mu$ candidate,
          if applicable.
     \end{itemize}

     We fit histograms of the \costhr and \sumecal signal and
     background MC distributions using polynomials of up to
     order eight to define the PDFs for those variables.
     The PDFs for the muon-PID quality level are normalized
     histograms, with one bin for each muon-PID quality level.

     For each of the eight signal $B$ decay modes, we construct a distinct $L_R$
     for each of the three $\tau$ channels ($e$, $\mu$, and $\pi$).
     This corresponds to 24 different likelihood ratios.
     In the final selection, described in section~\ref{sec:final-sel},
     we impose a minimum $L_R$ requirement for each $\tau$ channel
     in each of the eight \btohtl\ modes.

\begin{figure}[htb]
     \begin{center}
     \includegraphics[width=0.95\linewidth]{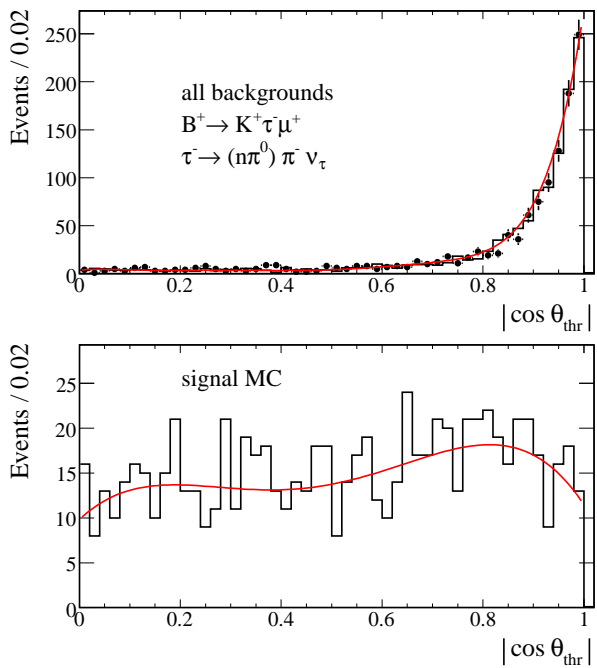}
     \caption{
         Distributions of \costhr for background (top)
         and signal MC (bottom), for the \btoktmp;
         $\tau^- \to (n\pi^0)\pi^-\nu_\tau$ channel.
         The points (solid line) in the top figure are the data (background MC).
         The background MC has been normalized to match the area of the data distribution.
         The normalization of the signal MC is arbitrary.
         The solid red curve is the result of the polynomial fit of the MC
         distribution.
     }
     \label{fig:costhr}
     \end{center}
   \end{figure}

     Figure~\ref{fig:costhr} shows the background and signal \costhr distributions for the $\pi$
     channel of the \btoktmp mode.
     The continuum background peaks sharply near $\costhr=1$ because the events have a back-to-back
     jet-like topology. 
     The signal \costhr distribution is roughly uniform because the detected decay products
     in \BB events are more isotropically distributed.

   %
   %----------------------------------------------------------
   %----------------------------------------------------------

\begin{figure*}
     \begin{center}
     \includegraphics[width=0.31\linewidth]{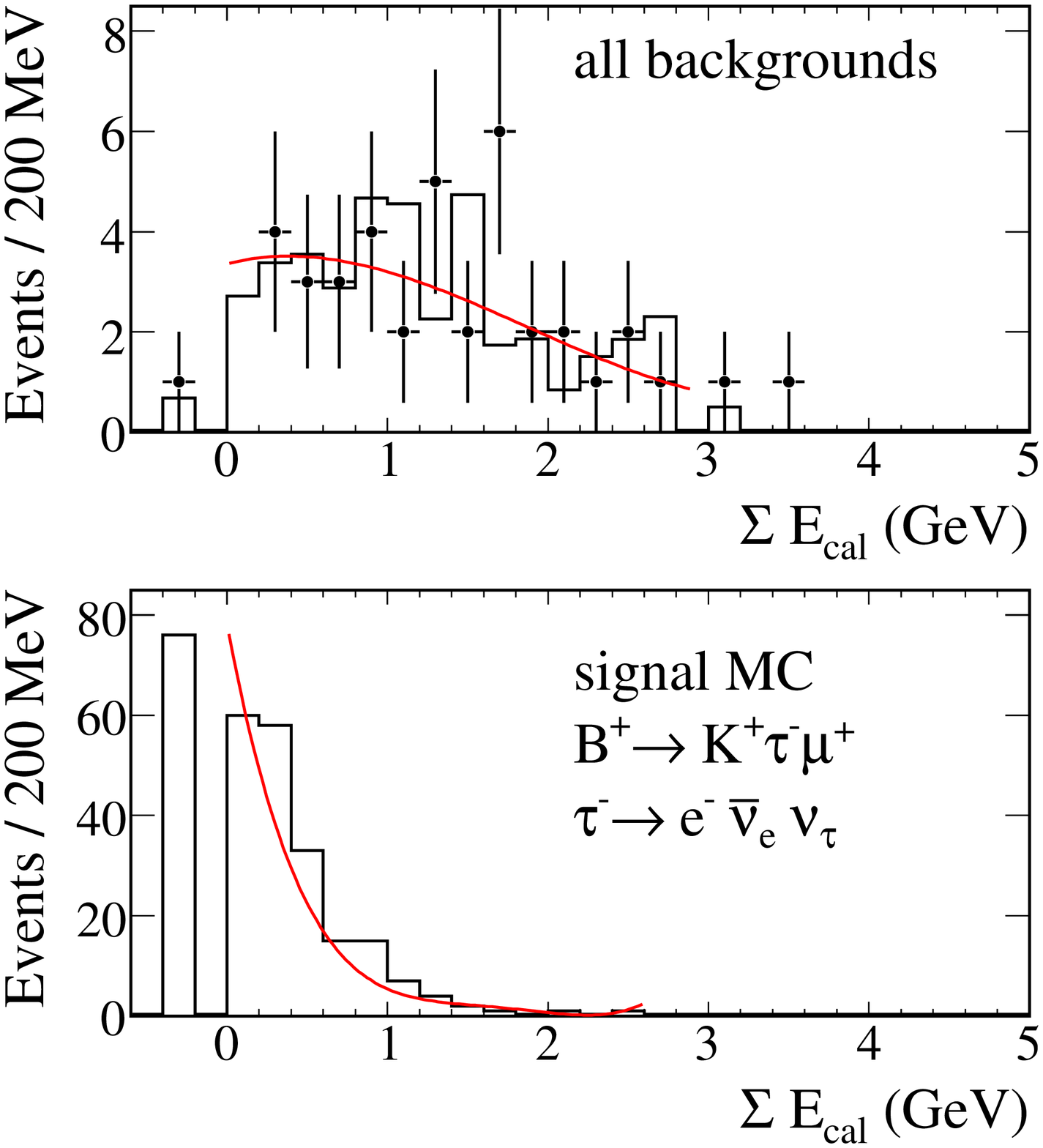}
     \includegraphics[width=0.31\linewidth]{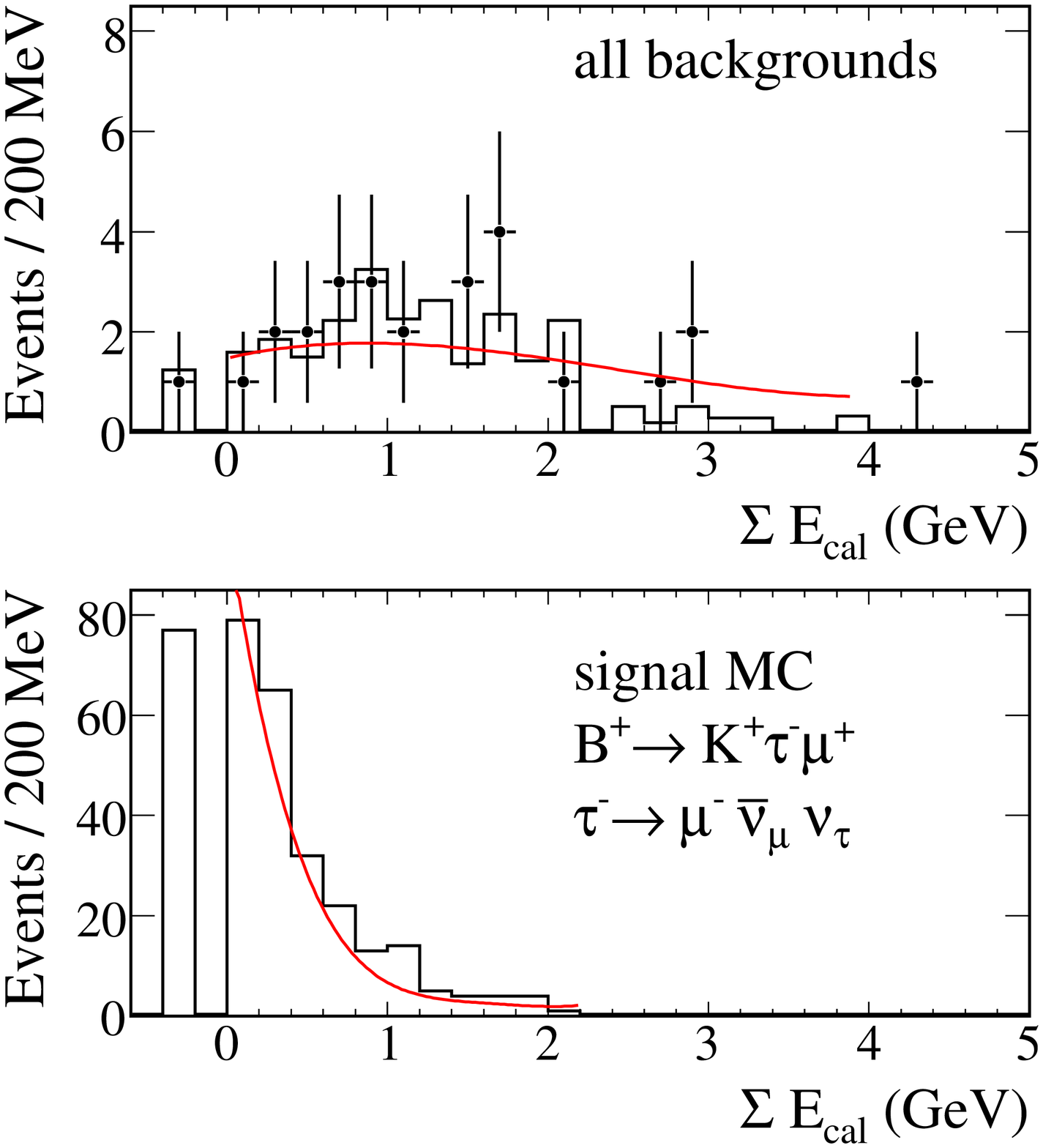}
     \includegraphics[width=0.31\linewidth]{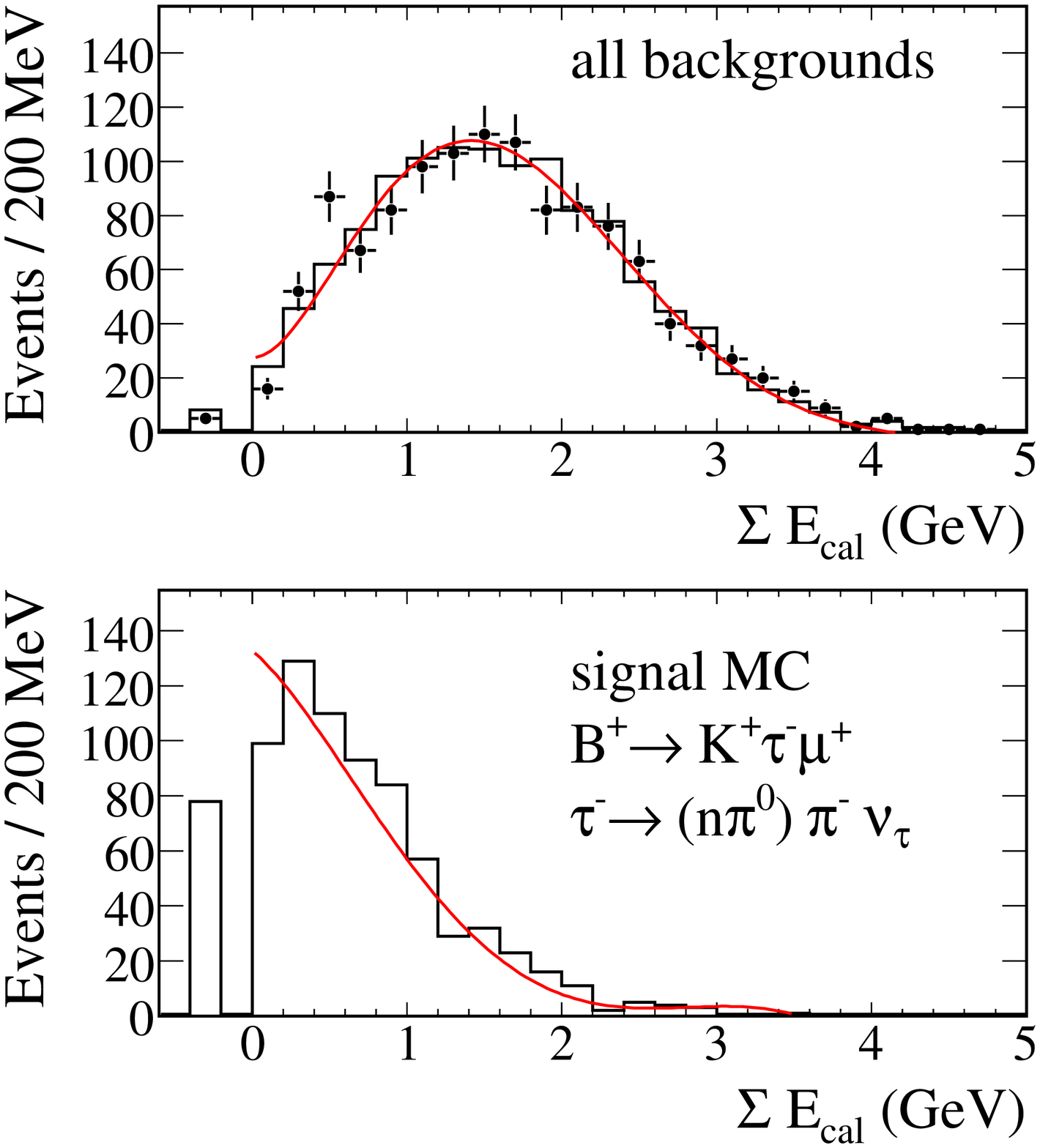}
     \caption{
         Distributions of \sumecal for background
         (top) and signal MC (bottom), for the \btoktmp mode;
         $\tau^- \to e^- \bar\nu_e \nu_\tau$ (left),
         $\tau^- \to \mu^- \bar\nu_\mu \nu_\tau$ (middle),
         and $\tau^- \to (n\pi^0)\pi^- \nu_\tau$ (right).
         The events where $\sumecal=0$ have been separated from the main distribution and
         plotted in a bin below zero for clarity.
         The points (solid line) in the top figure are the data (background MC).
         The background MC has been normalized to match the area of the data distribution.
         The normalization of the signal MC is arbitrary.
         The solid red curve is the result of the polynomial fit of the MC
         distribution.
     }
     \label{fig:sumcalE}
     \end{center}
   \end{figure*}

     Figure~\ref{fig:sumcalE} shows \sumecal distributions for the three $\tau$ channels
     of the \btoktmp mode.
     The events where $\sumecal=0$, due to the absence of unassociated neutral
     clusters above the minimum energy threshold, are not included in the polynomial
     fit and treated separately.
     The $\sumecal=0$ events are plotted below zero in Figure~\ref{fig:sumcalE} for clarity.
     The signal MC \sumecal distributions peak at zero, as expected, while the
     background rarely has $\sumecal=0$ but rather has a distribution that peaks between 1
     and 2 GeV.
     The signal MC \sumecal distributions for the $\pi$ channel extend to higher
     values, compared to the $e$ and $\mu$ channels, due to hadronic $\tau$
     decays that produce a single $\pi^\pm$ with one or more neutral pions.

   %
   %----------------------------------------------------------
   %----------------------------------------------------------

\begin{figure}[h]
     \begin{center}
     \includegraphics[width=0.95\linewidth]{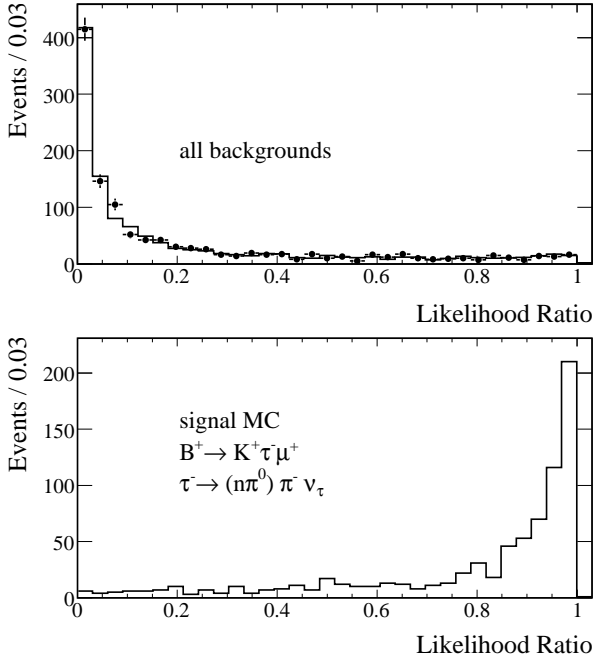}
     \caption{
         Likelihood ratio ($L_R$) output distributions of background (top)
         and signal MC (bottom), for the \btoktmp;
         $\tau^- \to (n\pi^0) \pi^-\nu_\tau$ channel.
         The points (solid line) in the top figure are the data (background MC).
         The background MC has been normalized to match the area of the data distribution.
         The normalization of the signal MC is arbitrary.
     }
     \label{fig:lhr}
     \end{center}
   \end{figure}

     Figure~\ref{fig:lhr} shows background
     and signal MC $L_R$ distributions for the \btoktmp\ ;
     $\tau^- \to (n\pi^0)\pi^-\nu_\tau$ channel.
     The background peaks sharply near zero and the signal peaks
     sharply near one.
     The value of the $L_R$ selection for each $\tau$ channel in each
     of the eight signal modes is chosen by determining the lowest
     upper limit on the branching fractions under the null hypothesis
     with MC pseudo-experiments.
     We vary the minimum $L_R$ requirement in intervals of 0.05.

   %----------------------------------------------------------
   %----------------------------------------------------------

   %%================================================================================================
     \section{ signal and background estimation }

     In our signal selection, we require the indirectly reconstructed $\tau$ mass
     $m_\tau$ to be within $\pm60$\mevcc of the world average $\tau$ mass
     1.777\gevcc~\cite{pdg}.
     The relative signal efficiency after the $m_\tau$ signal window requirement is around
     84\% (78\%) for the \btohtm\ (\btohte) modes.
     We optimized the $m_\tau$ signal windows, considering windows in the range
     of $\pm 50$~\mevcc to $\pm 175$~\mevcc.
     Our optimization metric was the average expected signal branching fraction 
     90\% confidence level upper limit from a set of toy experiments simulating
     background-only datasets.
        In each toy experiment, we generate a value for the observed number of events
        in the signal window using a random number that we take from a Poisson distribution
        with the mean value set to the expected number of background events.
     We find that a $m_\tau$ signal window of $\pm 60$~\mevcc gives the lowest expected
     branching fraction upper limits for all $\tau$ decay channels.

     The background distribution in $m_\tau$ is very wide and slowly varying.
     We use a broad $m_\tau$ sideband from 0 to 3.5\gevcc, excluding
     the signal window, to estimate the background
     in the $m_\tau$ signal window with
     \begin{equation}
       b \ = \ R_b \, N_{sb},
     \end{equation}
     where $b$ is the number of background events in the signal window, $N_{sb}$ is the number of
     background events in the $m_\tau$ sideband, and $R_b$ is the expected
     signal-to-sideband ratio ($b/N_{sb}$).
     The ratio $R_b$ is determined from the ratio of selected background events in the
     $m_\tau$ signal window ($b$) and the $m_\tau$ sideband ($N_{sb}$) in the
     background Monte Carlo.

     Figures~\ref{fig:mtau-ktaul} and~\ref{fig:mtau-pitaul} show the
     observed, signal MC, and background MC $m_\tau$ distributions for the \btoktl\ and \btoptl\ modes,
     respectively.
     Table~\ref{tab:bf-lim-results} gives the results for the observed numbers of sideband
     events $N_{sb,i}$,
     signal-to-sideband ratios $R_{b,i}$,
     expected numbers of background events $b_i$, numbers of observed
     events $n_i$, and signal efficiencies $\epsilon_{h\tau\ell,i}$
     for each $\tau$ channel $i$.
     All of the observed numbers of events $n_i$ in the $m_\tau$ signal window
     are statistically consistent with the expected backgrounds $b_i$, thus
     there is no evidence for any \btohtl\ decay.

   %%======================================================================================
     \section{ Systematic uncertainties }
     \label{sec:results}

     Since we normalize our \btohtl\ signals using the \btodlnu\ control sample,
     many systematic uncertainties cancel, such as the ones coming from
     the absolute \Btag efficiency uncertainty
     and the tracking efficiency uncertainty.
     We evaluate systematic uncertainties on the efficiency of the minimum
     $L_R$ requirement by varying the signal and background PDFs for each
     $L_R$.
         We use the \btodlnu control sample in place of the signal Monte
         Carlo as a variation of the signal \sumecal PDF.
         A uniform distribution is used in place of the nominal polynomial fit
         as the variation of the signal \costhr PDF.
         The efficiency for each lepton PID level is varied by $\pm 2.5$\% for
         the VL,L, and T levels and $\pm 3.2$\% for the VT level.
         The data $m_\tau$ sideband is used in place of the Monte Carlo as
         a variation of the background PDFs.

     Our largest sources of systematic uncertainty come from variations
     in modeling the data distributions of the \sumecal and \costhr $L_R$
     inputs when compared to the nominal background MC PDFs.
     The changes in $\epsilon_{h\tau\ell,i}$ from the variations are added
     in quadrature.
     We determine systematic uncertainties as high as 1.1\%, with 
     the largest ones coming from the
     \btoktem; $\tau^+ \to e^+\nu_{e}\bar{\nu_\tau}$ and
     $\tau^+ \to (n\pi^0)\pi^+\bar{\nu_\tau}$ channels.

     The \btoptl\ modes require $\pi$-PID, while the \btodlnu\ control sample
     requires $K$-PID.
     We evaluate a systematic uncertainty on
     $\epsilon_{\pi\tau\ell} / \epsilon_{D\ell\nu}$
     by measuring the $\pi$-PID and $K$-PID efficiencies using the
     \btodlnu\ control sample with and without the $K$-PID or $\pi$-PID
     requirements.
     The measured efficiencies in data are consistent with the MC simulation.
     Based on the results from the \btodmnu\ and \btodenu samples, we
     assign systematic uncertainties of 1.8\% and 1.0\% to
     $\epsilon_{\pi\tau\mu} / \epsilon_{D\mu\nu}$
     and
     $\epsilon_{\pi\tau e} / \epsilon_{D e \nu}$,
     respectively.

\begin{figure*}
     \begin{center}
     \includegraphics[width=0.49\linewidth]{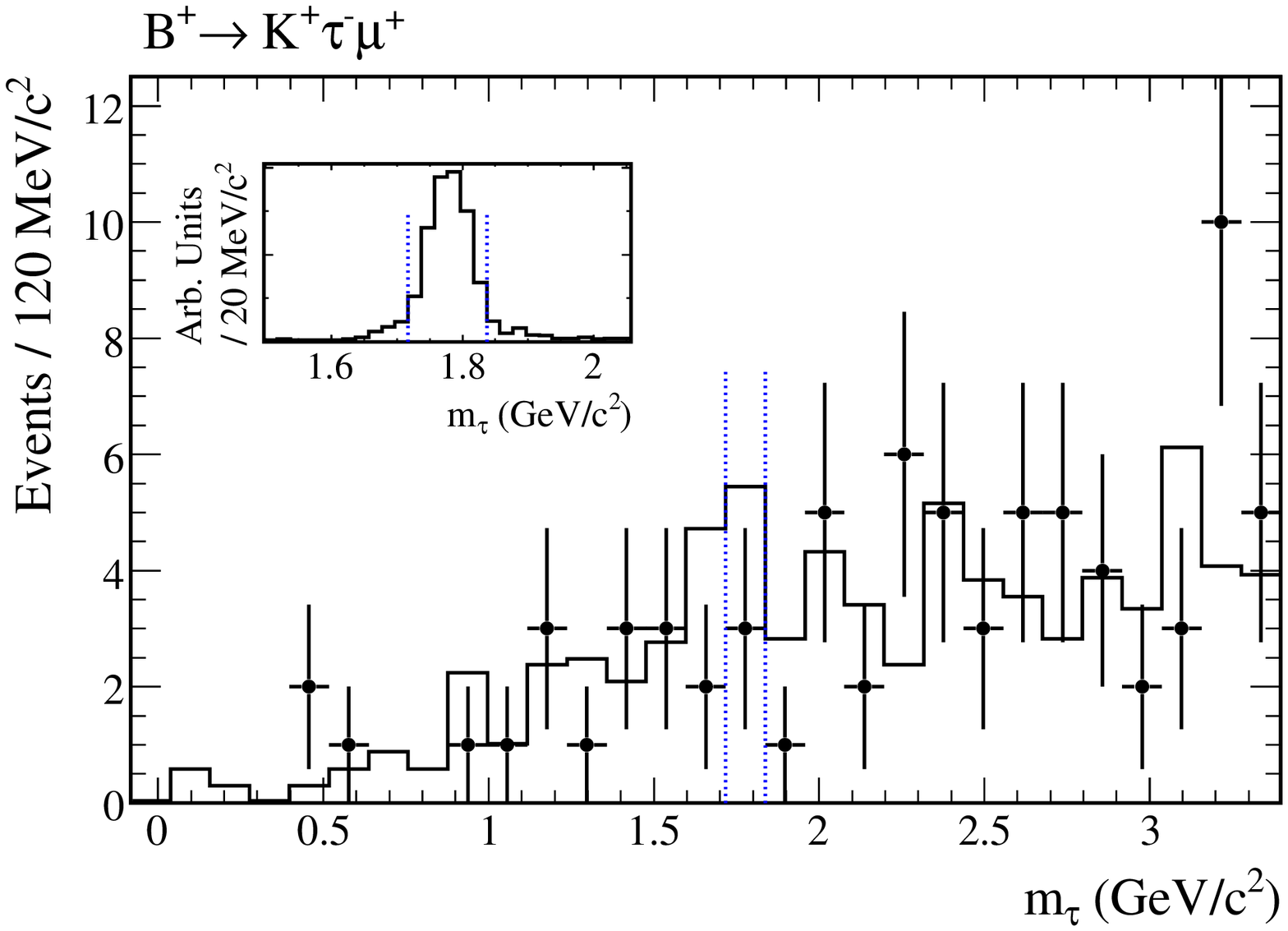}
     \includegraphics[width=0.49\linewidth]{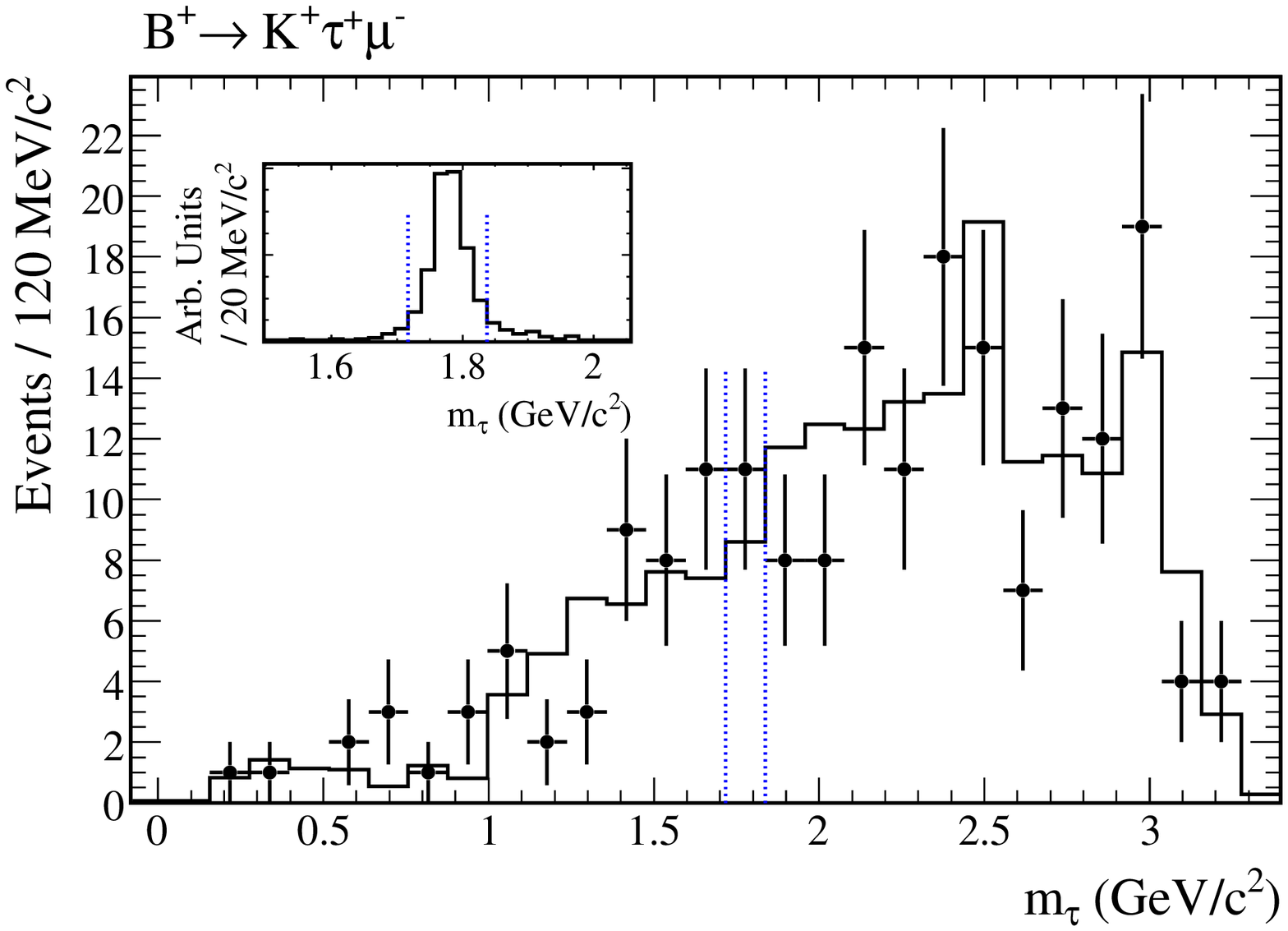}
     \includegraphics[width=0.49\linewidth]{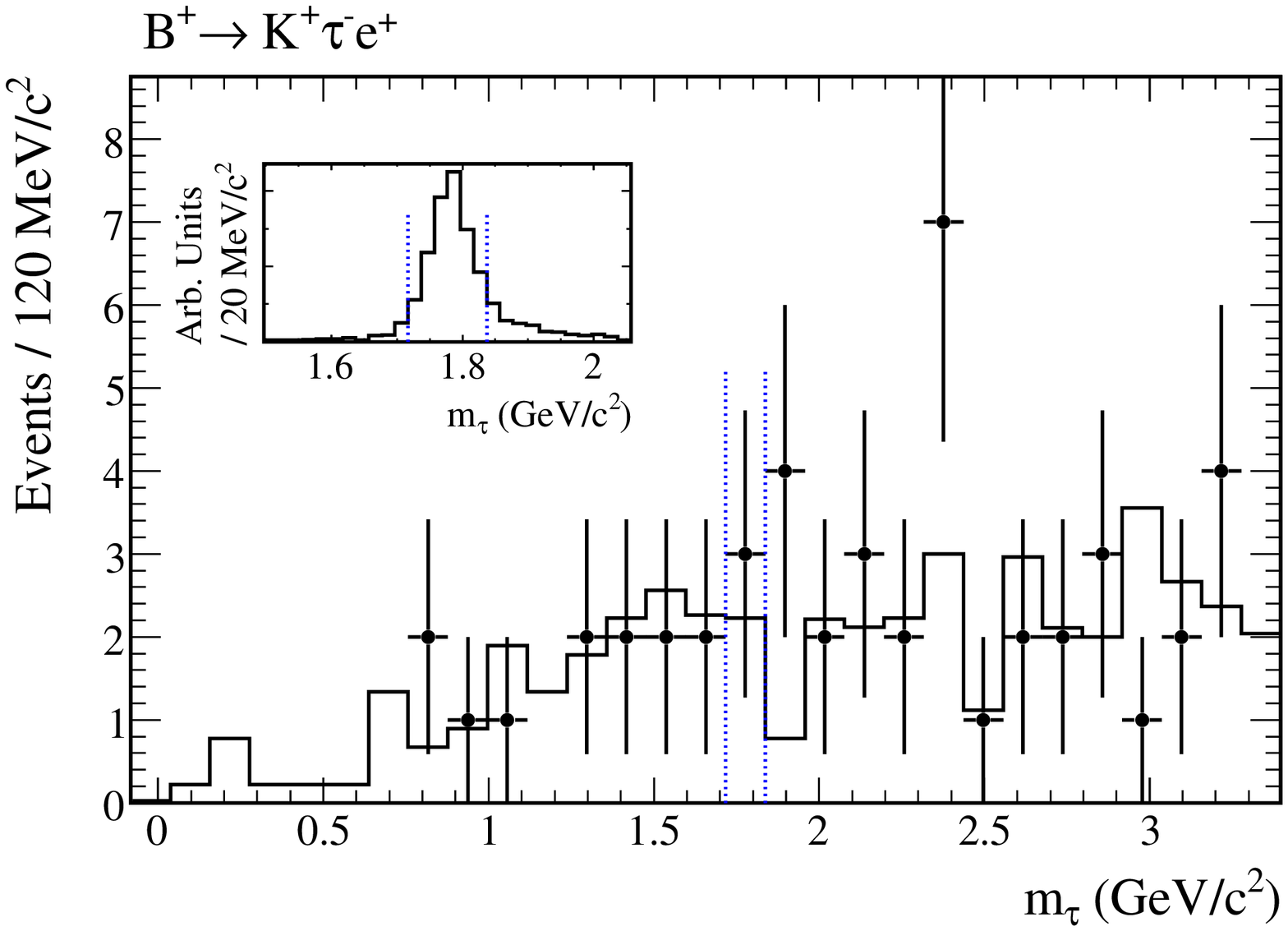}
     \includegraphics[width=0.49\linewidth]{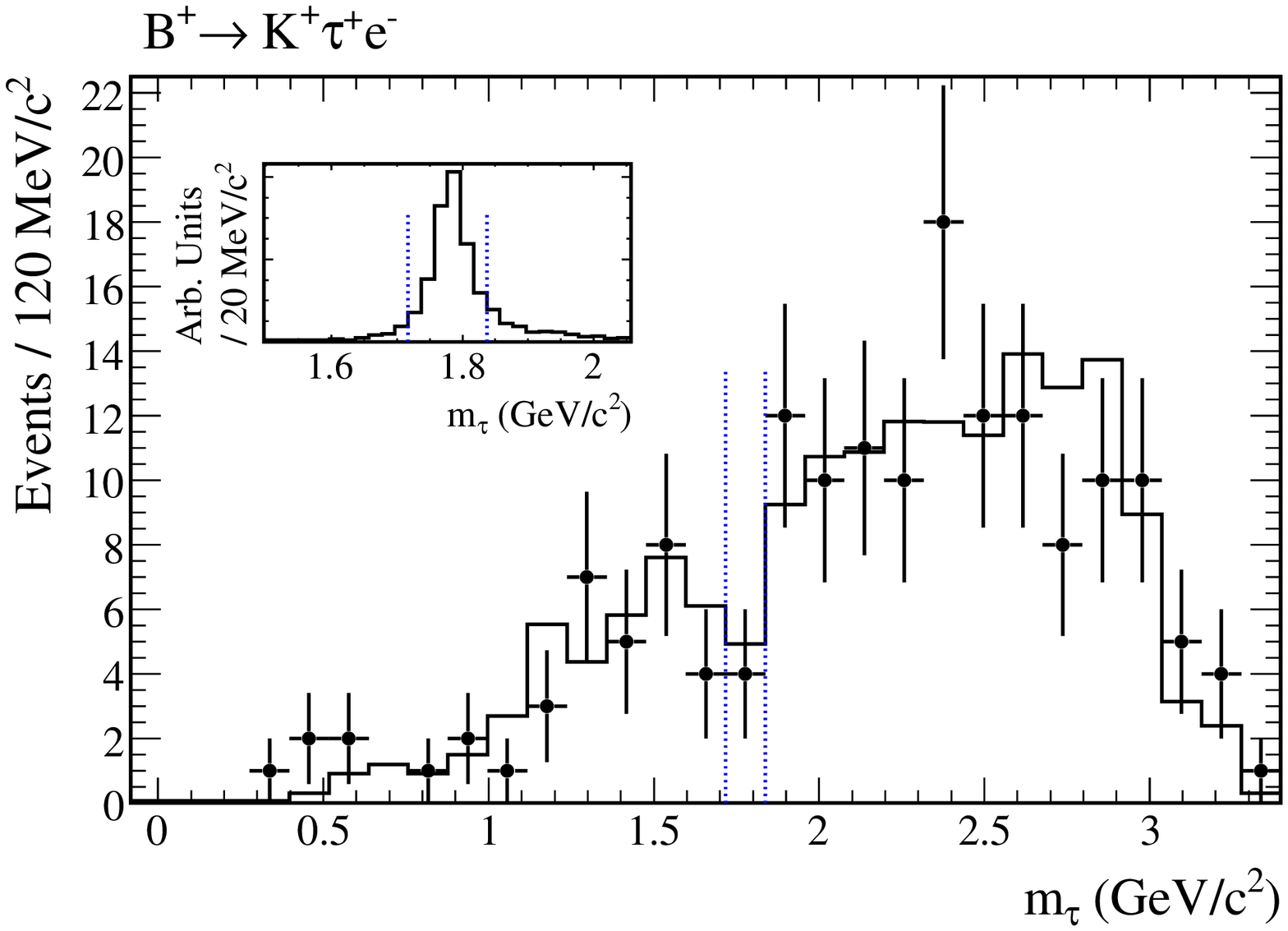}
     \caption{
         Observed distributions of the $\tau$ invariant mass for the \btoktl modes.
         The distributions show the sum of the three $\tau$ channels ($e$, $\mu$, $\pi$).
         The points with error bars are the data.
         The solid line is the background MC which has been normalized to the
         area of the data distribution.
         The dashed vertical lines indicate the $m_\tau$ signal window range.
         The inset shows the $m_\tau$ distribution for signal MC.
     }
     \label{fig:mtau-ktaul}
     \end{center}
   \end{figure*}

\begin{figure*}
     \begin{center}
     \includegraphics[width=0.49\linewidth]{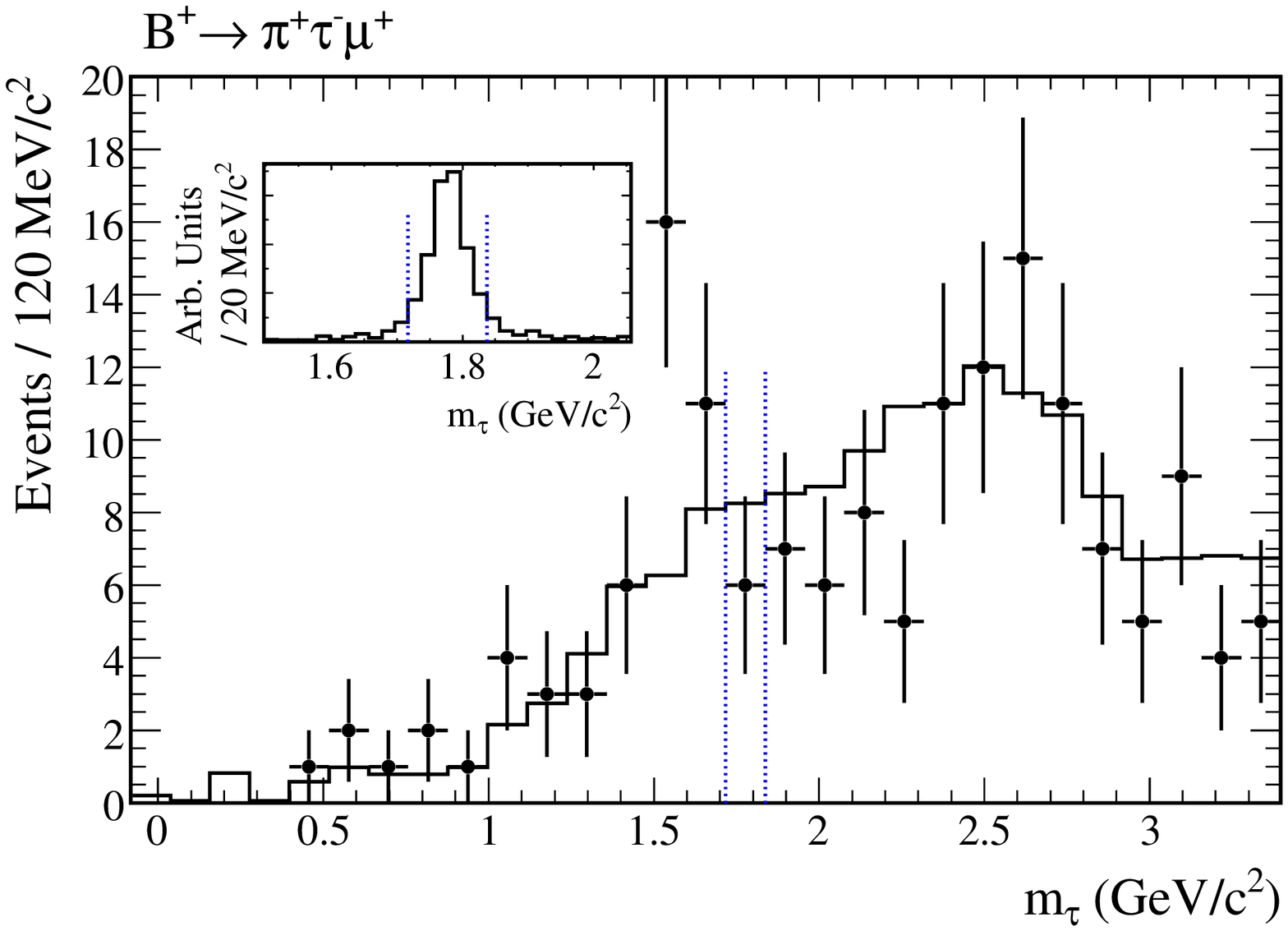}
     \includegraphics[width=0.49\linewidth]{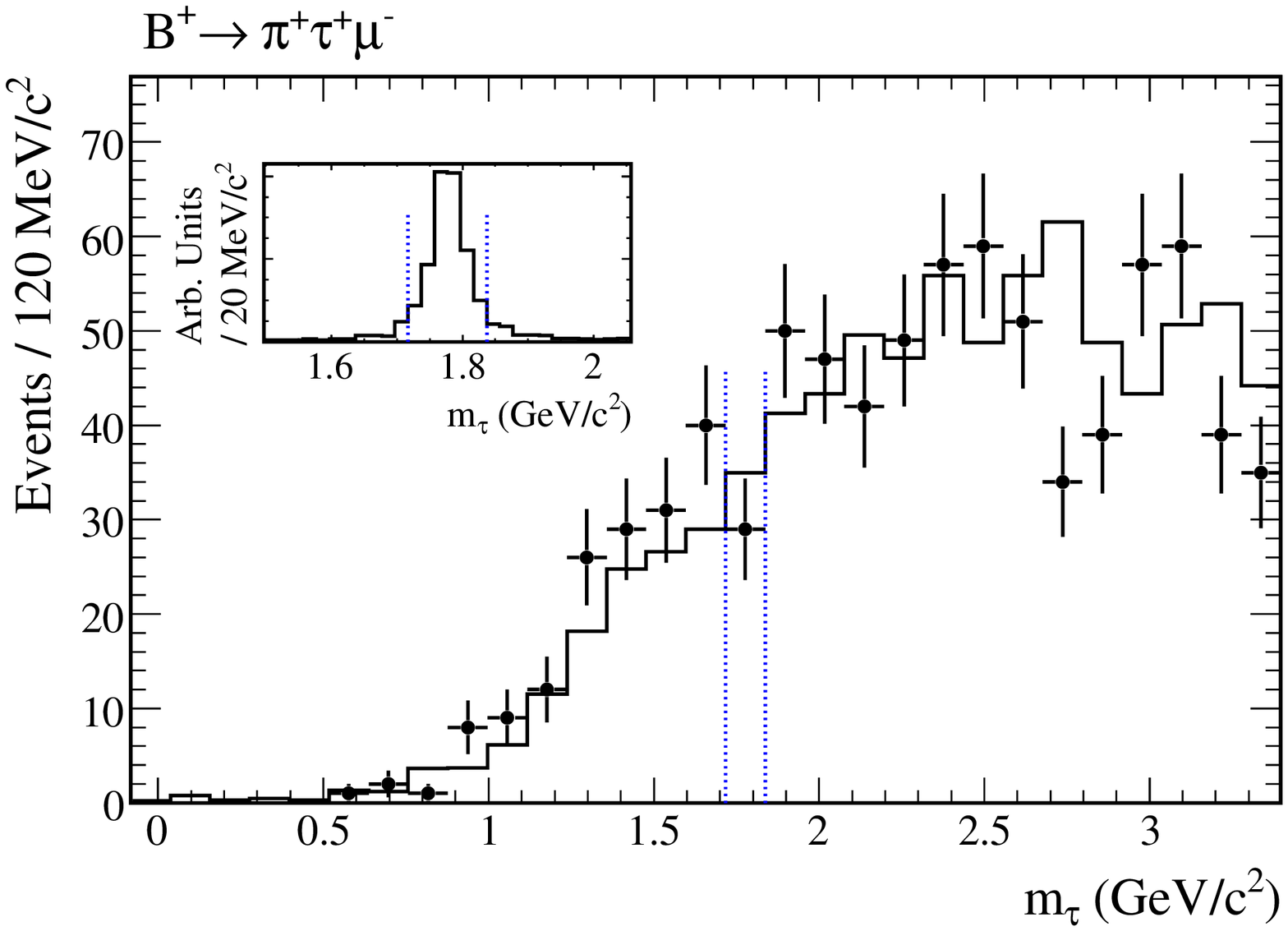}
     \includegraphics[width=0.49\linewidth]{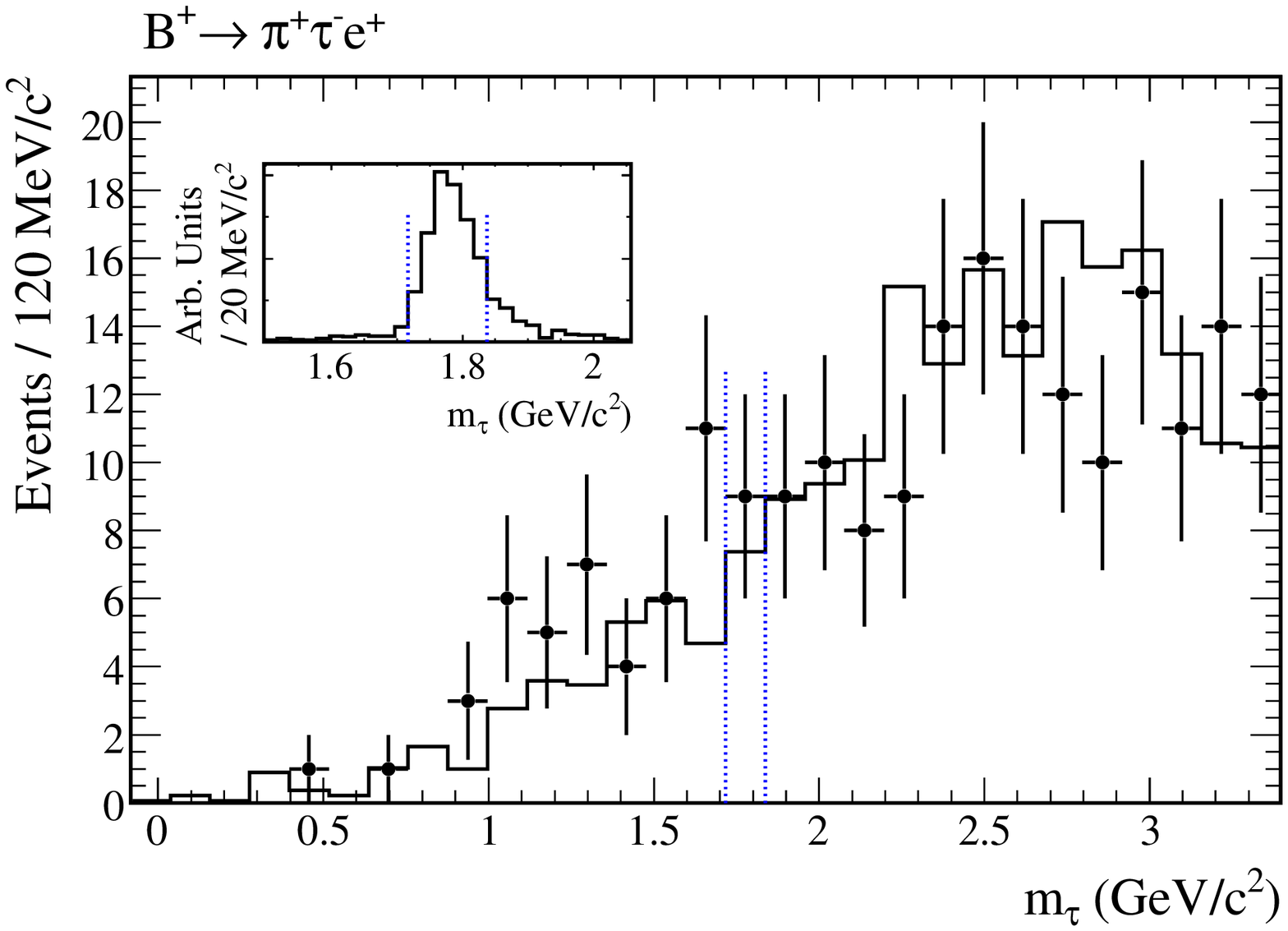}
     \includegraphics[width=0.49\linewidth]{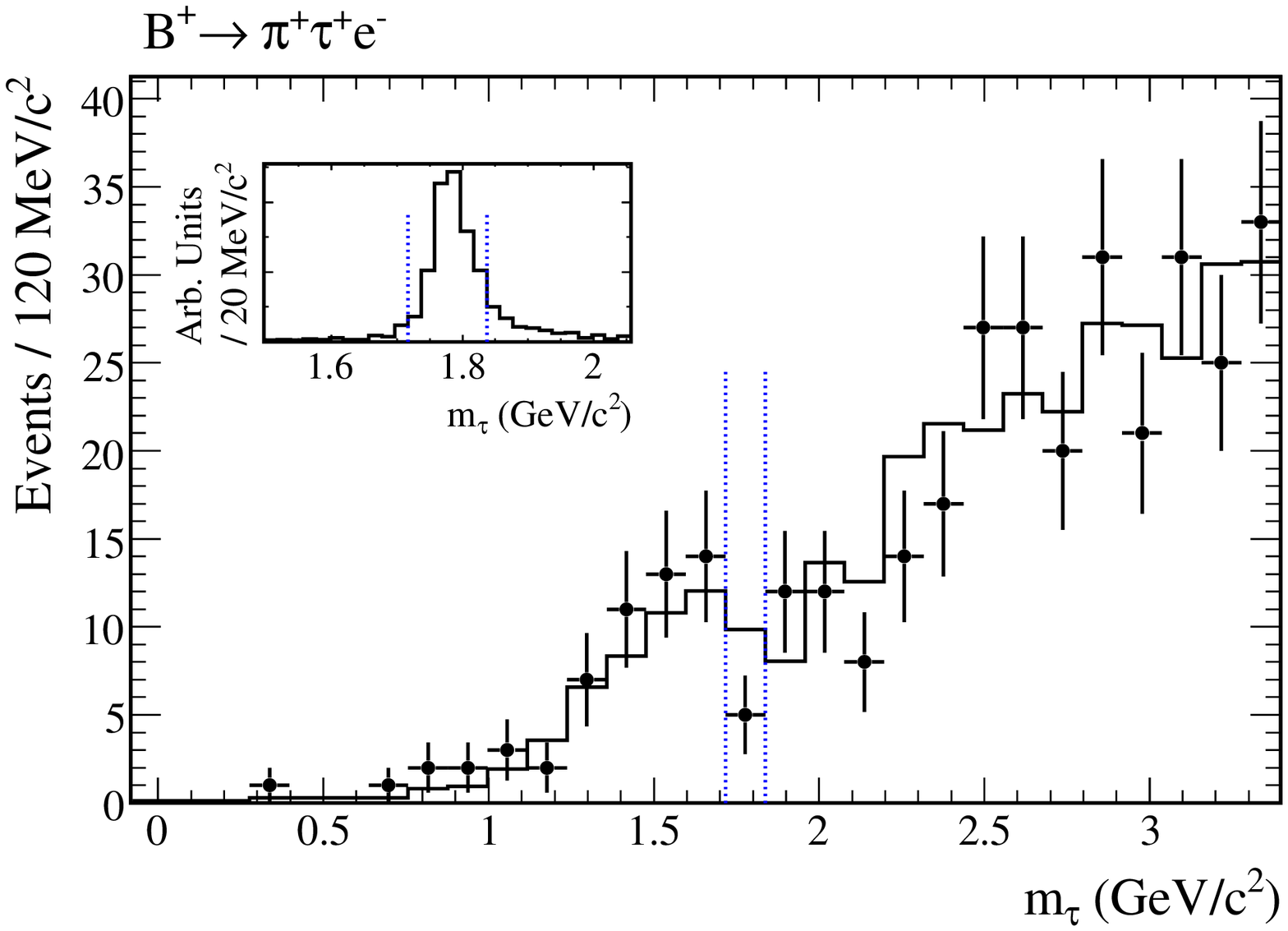}
     \caption{
         Observed distributions of the $\tau$ invariant mass for the \btoptl modes.
         The distributions show the sum of the three $\tau$ channels ($e$, $\mu$, $\pi$).
         The points with error bars are the data.
         The solid line is the background MC which has been normalized to the
         area of the data distribution.
         The dashed vertical lines indicate the $m_\tau$ signal window range.
         The inset shows the $m_\tau$ distribution for signal MC.
     }
     \label{fig:mtau-pitaul}
     \end{center}
   \end{figure*}

     The uncertainty on the signal-to-sideband ratio $R_b$ is the statistical
     uncertainty from the Monte Carlo sample used to determine its value.
     Figures~\ref{fig:mtau-ktaul} and~\ref{fig:mtau-pitaul} show good agreement between the Monte Carlo
     and observed data distributions in the sidebands.  No additional systematic
     uncertainty is included in $R_b$.

     The tag efficiency ratio
     $\epsilon^{h\tau\ell}_{\rm tag} / \epsilon^{D\ell\nu}_{\rm tag}$
     is evaluated using two independent Monte Carlo samples: one where
     the tag-side $B$ meson decays to all possible final states and
     another where the tag-side $B$ meson is forced to decay to most of
     the modes that comprise the tag-side reconstruction.
     The value of the ratio is taken from the first sample.
     The systematic error on the ratio is the difference in the ratio
     between the two samples.
     The overall uncertainty on the ratio, given in Table~\ref{tab:tagside-eff-ratio}, is
     the sum in quadrature of the statistical and systematic uncertainties
     on the ratio.

   %%================================================================================================

     \section{ Branching fraction results}
     \label{sec:final-sel}

     We determine the branching fraction for each of the eight \btohtl
     modes using a likelihood function which is the product of
     three Poisson PDFs, one for each of the three $\tau$ channels.
     The expected number of events in a particular $\tau$ channel
     is given by
     \begin{eqnarray}
       n_i & = & N_{h\tau\ell,i} + b_i \\
       n_i & = & {\cal B}_{h\tau\ell} \ \epsilon_{h\tau\ell,i} \ S_0  +  b_i,
     \end{eqnarray}
     where $N_{h\tau\ell,i}$ ($b_i$) is the expected number of signal (background) events
     in channel $i$.
     Total uncertainties on the signal efficiency $\epsilon_{h\tau\ell,i}$,
     common factor $S_0$,
     and expected background $b_i$
     are included by convolving the likelihood with Gaussian distributions
     in $\epsilon_{h\tau\ell,i}$, $S_0$, and $b_i$.

     We set 90\% confidence intervals on the branching fractions of
     the eight \btohtl\ modes assuming uniform three-body phase space decays
     using the likelihood ratio ordering principle of Feldman and
     Cousins~\cite{ref:feldman-cousins} to construct the confidence belts.

      The 90\% C.L. upper limits on the \btohtl\ branching fractions are between
      $1.5 \times 10^{-5}$ and $7.4 \times 10^{-5}$.
      Table~\ref{tab:bf-lim-results} includes the final results for the \btohtl
      branching fraction and 90\% C.L. upper limits.
      In Table~\ref{tab:combined-6chan-limits}, we give combined results for
       ${\cal B}(B^+ \to h^+ \tau \ell) \equiv {\cal B}(B^+ \to h^+ \tau^- \ell^+) + {\cal B}(B^+ \to h^+ \tau^+ \ell^-)$
      with the assumption ${\cal B}(B^+ \to h^+ \tau^- \ell^+) = {\cal B}(B^+ \to h^+ \tau^+ \ell^-)$.

     In the analysis of Black, Han, He, and Sher~\cite{black},
     the $B\to K\tau\mu$ and $B\to \pi\tau\mu$ branching fractions
     are proportional to $\Lambda^{-4}_{\bar b s}$ and
     $\Lambda^{-4}_{\bar b d}$, which are the new physics energy scales for
     the corresponding fermionic effective operators for these decays.
     Using the limits ${\cal B}(B^+ \to \pi^+ \tau \mu)<7.2\times 10^{-5}$
     and ${\cal B}(B^+ \to K^+ \tau \mu)<4.8\times 10^{-5}$,
     we improve the model-independent bounds on the energy scale of
     new physics in flavor-changing operators reported in~\cite{black}
     from
     $\Lambda_{\bar b d}>2.2$~TeV and $\Lambda_{\bar b s}>2.6$~TeV
     to
     $\Lambda_{\bar b d}>11$~TeV and $\Lambda_{\bar b s}>15$~TeV.

   %
   %----------------------------------------------------------
   %----------------------------------------------------------
   %
   %----------------------------------------------------------
   %----------------------------------------------------------

     %%---------------------
     \begin{table*}[]
       \begin{center}
         \caption{ Results for the observed sideband events $N_{sb,i}$,
                   signal-to-sideband ratio $R_{b,i}$,
                   expected background events $b_i$, number of observed
                   events $n_i$, signal efficiency $\epsilon_{h\tau\ell,i}$
                   (assuming uniform three-body phase space decays)
                   for each $\tau$ channel $i$ and
                   \btohtl~\cite{chargecon} branching fraction
                   central value and
                   90\% C.L. upper limits (UL).
                   All uncertainties include statistical and systematic sources.
                  }
         \label{tab:bf-lim-results}
         \begin{tabular}{ccccccccc}
           \hline \hline
                 &
                 &
                 &
                 &
                 &
                 &
                 &  \multicolumn{2}{c}{${\cal B}(\btohtl)$ $(\times 10^{-5})$}  \smtvs \\
           Mode  &  $\tau$ channel
                 & \ \ $N_{sb,i}$ \ \
                 & \ \ \ \ \ \ \ \ $R_{b,i}$ \ \ \ \ \ \ \ \
                 &  \ \ \ \ \ $b_i$ \ \ \ \ \
                 &  \ \ \ $n_i$ \ \ \
                 &  $\epsilon_{h\tau\ell,i}$
                 &  \ \ central value \ \
                 &  \ \ 90\% C.L. UL \ \ \smtvs \\
            \hline \hline
                 &  $e$    &  22  &   $0.02 \pm 0.01$   &  $0.4 \pm 0.2$  &  2  &  $(2.6 \pm 0.2)$\%  &  \\
      \btoktmp   &  $\mu$  &   4  &   $0.08 \pm 0.05$   &  $0.3 \pm 0.2$  &  0  &  $(3.2 \pm 0.4)$\%  & \ \ $0.8 \ ^{+1.9}_{-1.4}$ \ & \ $<4.5$ \ \   \\
                 &  $\pi$  &  39  &  $0.045 \pm 0.020$  &  $1.8 \pm 0.8$  &  1  &  $(4.1 \pm 0.4)$\%  &  \\
            \hline
                 &  $e$    &   5  &   $0.03 \pm 0.01$   &  $0.2 \pm 0.1$  &  0  &  $(3.7 \pm 0.3)$\%  &  \\
      \btoktmm   &  $\mu$  &   3  &   $0.06 \pm 0.03$   &  $0.2 \pm 0.1$  &  0  &  $(3.6 \pm 0.7)$\%  & \ \ $-0.4 \ ^{+1.4}_{-0.9}$ \ & \ $<2.8$ \ \  \\
                 &  $\pi$  & 153  &  $0.045 \pm 0.010$  &  $6.9 \pm 1.5$  & 11  &  $(9.1 \pm 0.5)$\%  &  \\
            \hline \hline
                 &  $e$    &   6  &  $0.095 \pm 0.020$  &  $0.6 \pm 0.1$  &  2  &  $(2.2 \pm 0.2)$\%  &  \\
      \btoktep   &  $\mu$  &   4  &  $0.025 \pm 0.010$  &  $0.1 \pm 0.1$  &  0  &  $(2.7 \pm 0.6)$\%  & \ \ $0.2 \ ^{+2.1}_{-1.0}$ \ & \ $<4.3$ \ \  \\
                 &  $\pi$  &  33  &  $0.045 \pm 0.015$  &  $1.5 \pm 0.5$  &  1  &  $(4.8 \pm 0.6)$\%  &  \\
            \hline
                 &  $e$    &   8  &   $0.10 \pm 0.06$   &  $0.8 \pm 0.5$  &  0  &  $(2.8  \pm 1.1)$\%   &  \\
      \btoktem   &  $\mu$  &   3  &  $0.045 \pm 0.020$  &  $0.1 \pm 0.1$  &  0  &  $(3.2 \pm 0.7)$\%  & \ \ $-1.3 \ ^{+1.5}_{-1.8}$ \ & \ $<1.5$ \ \  \\
                 &  $\pi$  & 132  &  $0.035 \pm 0.010$  &  $4.6 \pm 1.3$  &  4  &  $(8.7  \pm 1.2)$\%   &  \\
            \hline \hline
                 &  $e$    &  55  &  $0.017 \pm 0.010$  &  $0.9 \pm 0.6$  &  0  &  $(2.3 \pm 0.2)$\%  &  \\
      \btoptmp   &  $\mu$  &  10  &   $0.11 \pm 0.04$   &  $1.1 \pm 0.4$  &  2  &  $(2.9 \pm 0.4)$\%  & \ \ $0.4 \ ^{+3.1}_{-2.2}$ \ & \ $<6.2$ \ \  \\
                 &  $\pi$  &  93  &  $0.035 \pm 0.010$  &  $3.3 \pm 0.9$  &  4  &  $(2.8 \pm 0.2)$\%  &  \\
            \hline
                 &  $e$    & 171  &  $0.012 \pm 0.003$  &  $2.1 \pm 0.5$  &  2  &  $(3.8 \pm 0.3)$\%  &  \\
      \btoptmm   &  $\mu$  &  89  &   $0.04 \pm 0.01$   &  $3.6 \pm 0.9$  &  4  &  $(4.8 \pm 0.3)$\%  & \ \ $0.0 \ ^{+2.6}_{-2.0}$ \ & \ $<4.5$ \ \  \\
                 &  $\pi$  & 512  &  $0.050 \pm 0.005$  &  $25  \pm   3$  & 23  &  $(9.1 \pm 0.6)$\%  &  \\
            \hline \hline
                 &  $e$    &   1  &  $0.050 \pm 0.025$  &  $0.1 \pm 0.1$  &  1  &  $(2.0 \pm 0.8)$\%  &  \\
      \btoptep   &  $\mu$  &  16  &  $0.025 \pm 0.010$  &  $0.4 \pm 0.2$  &  1  &  $(2.8 \pm 0.3)$\%  & \ \ $2.8 \ ^{+2.4}_{-1.9}$ \ & \ $<7.4$ \ \  \\
                 &  $\pi$  & 172  &  $0.035 \pm 0.008$  &  $6.0 \pm 1.4$  &  7  &  $(5.8 \pm 0.3)$\%  &  \\
            \hline
                 &  $e$    &  31  &  $0.033 \pm 0.013$  &  $1.0 \pm 0.4$  &  0  &  $(2.9 \pm 0.3)$\%  &  \\
      \btoptem   &  $\mu$  & 247  &  $0.012 \pm 0.005$  &  $3.0 \pm 1.2$  &  2  &  $(4.6 \pm 0.4)$\%  & \ \ $-3.1 \ ^{+2.4}_{-2.1}$ \ & \ $<2.0$ \ \  \\
                 &  $\pi$  &  82  &   $0.07 \pm 0.03$   &  $5.7 \pm 2.5$  &  3  &  $(3.7 \pm 1.0)$\%  &  \\
            \hline \hline
         \end{tabular}
       \end{center}
     \end{table*}

   \begin{table}
     \begin{center}
       \caption{
          Branching fraction central values and 90\% C.L. upper limits (UL)
          for the combination
          ${\cal B}(B^+ \to h^+ \tau \ell) \equiv {\cal B}(B^+ \to h^+ \tau^- \ell^+) + {\cal B}(B^+ \to h^+ \tau^+ \ell^-)$
          with the assumption ${\cal B}(B^+ \to h^+ \tau^- \ell^+) = {\cal B}(B^+ \to h^+ \tau^+ \ell^-)$.
          The uncertainties include statistical and systematic sources.
       }
       \label{tab:combined-6chan-limits}
       \begin{tabular}{lcc}
         \hline \hline
                      &  \multicolumn{2}{c}{ ${\cal B}(B \to h \tau \ell)$ $(\times 10^{-5})$} \\
         \multicolumn{1}{c}{ \ \ \ \ \  Mode \ \ \ \ \ \ \ \ \ \ \ }       &  \ \ \ \ central value\ \ \ \ & \ \ \ 90\% C.L. UL \ \ \ \\
         \hline \hline
         $B^+ \to K^+ \tau \mu$   &  $0.0\ ^{+2.7}_{-1.4}$   &  $<4.8$   \\
         \hline
         $B^+ \to K^+ \tau e$     &  $-0.6\ ^{+1.7}_{-1.4}$  &  $<3.0$   \\
         \hline
         $B^+ \to \pi^+ \tau \mu$ &  $0.5\ ^{+3.8}_{-3.2}$  &  $<7.2$   \\
         \hline
         $B^+ \to \pi^+ \tau e$   &  $2.3\ ^{+2.8}_{-1.7}$  &  $<7.5$   \\
         \hline \hline
       \end{tabular}
     \end{center}
   \end{table}

   %%======================================================================================
     \section{ Summary and conclusions }

     We have searched for the lepton flavor violating decays \btohtl.
     We find no evidence for these decays and set 90\% C.L. upper limits on the
     branching fractions of a few times $10^{-5}$.
     The results for the \btoktm mode supersede our previous result~\cite{babar-ktm}.
     The results for \btokte, \btoptm, \btopte\ modes are the first experimental
     limits for these decays.
     We use our results to improve model-independent limits on the energy scale
     of new physics in flavor-changing operators~\cite{black} to
     $\Lambda_{\bar b d}>11$~TeV and $\Lambda_{\bar b s}>15$~TeV.

   %%%%%%%%%%%%%%%%%%%%%%%%%%%%%%%%%%%%%%%%%%%%%%%%%%%%%%%%%%%%%%%%%%%%%%%%%%%%%%%%%%%%%%%%%%%%

   %%======================================================================================
     \section*{ Acknowledgments }

     We are grateful for the 
extraordinary contributions of our \pep2\ colleagues in
achieving the excellent luminosity and machine conditions
that have made this work possible.
The success of this project also relies critically on the 
expertise and dedication of the computing organizations that 
support \babar.
The collaborating institutions wish to thank 
SLAC for its support and the kind hospitality extended to them. 
This work is supported by the
US Department of Energy
and National Science Foundation, the
Natural Sciences and Engineering Research Council (Canada),
the Commissariat \`a l'Energie Atomique and
Institut National de Physique Nucl\'eaire et de Physique des Particules
(France), the
Bundesministerium f\"ur Bildung und Forschung and
Deutsche Forschungsgemeinschaft
(Germany), the
Istituto Nazionale di Fisica Nucleare (Italy),
the Foundation for Fundamental Research on Matter (The Netherlands),
the Research Council of Norway, the
Ministry of Education and Science of the Russian Federation, 
Ministerio de Ciencia e Innovaci\'on (Spain), and the
Science and Technology Facilities Council (United Kingdom).
Individuals have received support from 
the Marie-Curie IEF program (European Union) and the A. P. Sloan Foundation (USA).

   %%%%%%%%%%%%%%%%%%%%%%%%%%%%%%%%%%%%%%%%%%%%%%%%%%%%%%%%%%%%%%%%%%%%%%%%%%%%%%%%%%%%%%%%%%%%


\begin{thebibliography}{99}
\bibitem{chargecon} Charge-conjugate decays are implied throughout the paper.
\bibitem{cmframe} All energies and momenta are in the \epem\ center-of-mass reference
            frame unless explicitly stated otherwise.
\end{thebibliography}
   \end{document}